\documentclass{iopart}

\usepackage{epsfig}
\usepackage{graphicx}
\usepackage{graphics}
\usepackage{dcolumn}
\usepackage{bm}
\usepackage{hyperref}
\usepackage{iopams}
\usepackage{setstack}
\usepackage{amsthm}

\begin{document}

\title{Frankenstein's Glue: \\Transition functions for approximate solutions}

\author{Nicol\'as  Yunes}

\address{Institute for Gravitational Physics and Geometry,
         Center for Gravitational Wave Physics,
         Department of Physics, The Pennsylvania State University,
         University Park, PA 16802-6300}

\eads{yunes@gravity.psu.edu}


\submitto{\CQG} 

%
\newcommand\be{\begin{equation}}
\newcommand\ba{\begin{eqnarray}}

\newcommand\ee{\end{equation}}
\newcommand\ea{\end{eqnarray}}
\newcommand\p{{\partial}}
\newcommand\remove{{{\bf{THIS FIG. OR EQS. COULD BE REMOVED}}}}

\newtheorem{theorem}{Theorem}

%

\begin{abstract}
  
  Approximations are commonly employed to find approximate solutions
  to the Einstein equations. These solutions, however, are usually
  only valid in some specific spacetime region. A global solution can
  be constructed by gluing approximate solutions together, but this
  procedure is difficult because discontinuities can arise, leading to
  large violations of the Einstein equations. In this paper, we
  provide an attempt to formalize this gluing scheme by studying
  transition functions that join approximate analytic solutions
  together. In particular, we propose certain sufficient conditions on
  these functions and proof that these conditions guarantee that the
  joined solution still satisfies the Einstein equations analytically
  to the same order as the approximate ones. An example is also
  provided for a binary system of non-spinning black holes, where the
  approximate solutions are taken to be given by a post-Newtonian
  expansion and a perturbed Schwarzschild solution. For this specific
  case, we show that if the transition functions satisfy the proposed
  conditions, the joined solution does not contain any violations to
  the Einstein equations larger than those already inherent in the
  approximations.  We further show that if these functions violate the
  proposed conditions, then the matter content of the spacetime is
  modified by the introduction of a matter-shell, whose stress-energy
  tensor depends on derivatives of these functions.
  
\end{abstract}

\maketitle

\section{Introduction}

The detection and modeling of gravitational radiation is currently one
of the primary driving forces in classical general relativity due to
the advent of
ground-based~\cite{Abramovici:1992ah,Willke:2002bs,Giazotto:1988gw,Ando:2002bv}
and space-born detectors~\cite{Danzmann:1997hm,LISA}. This radiation
is generated in highly dynamical spacetimes, whose exact metric has
not yet been found. Approximate solutions to the Einstein equations,
both analytical (such as post-Newtonian
solutions~\cite{Blanchet:2002av}) and numerical, have served to
provide insight on the dynamics of such spacetimes and the character
of the radiation produced. These approximations, however, have
inherent uncontrolled remainders, or errors, due to truncation of
certain higher order terms in the analytical case, or discretization
error in the numerical case.  An approximate global solution, then,
could be constructed by joining several approximate solutions together
in some overlap region~\cite{Yunes:2005nn,Yunes:2006iw}.

Before proceeding, we must distinguish between two different kinds of
joined solutions: mixed ones, where one joins an analytical solution
to a numerical one; and pure ones, where one joins two analytical
solutions that have different but overlapping regions of validity. A
special kind of mixed joined solutions have been created in the
context of the effective-one-body
formalism~\cite{Buonanno:2000ef,Buonanno:2005xu}, with the motivation
of providing accurate waveform templates to gravitational wave
interferometers. Pure joined solutions have been discussed in the
context of asymptotic matching, where one of the motivations is to use
the joined solution as initial data for relativistic
simulations~\cite{Yunes:2005nn,Yunes:2006iw}. In this paper, we
concentrate on pure joined solutions, although the methods and
conditions we find can be straightforwardly extended to mixed joined
solutions.

The construction of pure joined solutions is not always simple because
approximation methods tend to break down in highly dynamical
spacetimes. The main difficulty lies in that approximate solutions
usually depend on the existence of a background about which to perturb
the solution. However, in highly dynamical spacetimes, such a
background cannot usually be constructed.  Thus, in those scenarios,
the region of validity of the different approximations tends not to
overlap. In slightly less dynamical cases, the regions of validity can
overlap, but the different approximate metrics usually describe the
spacetime in different coordinates and parameters.
Patching~\cite{Bender} could be used to relate the different
approximate solutions, but this method usually leads to an
overdetermined system if we require the patched metric to be
differentiable at the junction. A better alternative is to relate the
approximate solutions via asymptotic matching, which guarantees that
adjacent metrics and their derivatives be asymptotic to each other in
some overlap region.

Asymptotic matching was developed as a technique of multiple-scale
analysis to solve non-linear partial differential
equations~\cite{Bender,Kevorkian}. In general relativity, this method
was first studied
in~\cite{Burke,Burke-Thorne,Death:1974o,Death:1975,Thorne:1984mz} and
it has recently had important applications to post-Newtonian
theory~\cite{Blanchet:2002av}, black hole perturbation
theory~\cite{Mino:1997bw} and initial data
constructions~\cite{Yunes:2005nn,Yunes:2006iw}. Asymptotic matching
requires that we compare the asymptotic expansions of the approximate
solutions inside the region of the manifold where the regions of
validity of the approximations overlap (the buffer zone.)  This method
then provides a coordinate and parameter transformation that relates
the approximate metrics, such that adjacent metrics and their
derivatives become asymptotic to each other inside the buffer zone. In
essence, asymptotically matched approximations are guaranteed to
represent the same metric components in the same coordinate system
inside the buffer zone.
 
After matching has been carried out, there is still freedom as to how
to join the matched approximate solutions together. The simplest way
to do so is through a weighted linear combination of approximate
solutions with transition functions.  This method was first considered
in depth in~\cite{Yunes:2005nn,Yunes:2006iw}, where a binary black
hole system was studied to construct initial data for numerical
simulations (the so-called Frankenstein approach). In that work, only
broad comments were made as to the type of allowed transition
functions, requiring only that the functions be ``differentiable
enough,'' and the the properties and conditions these functions must
satisfy were not studied. A priori, it might not be clear which
functions are allowed such that the global metric still approximately
solves the Einstein equations. For instance, it might seem natural to
use Heaviside functions to join the metrics together at a
hypersurface, as is done in the standard junction conditions of
general
relativity~\cite{Lanczos:1922,Lanczos:1924,Darmois:1927,Lichnerowicz,Misner:1964,Barrabes:1991ng,Israel:1966nc,MTW,poisson}.
These joining procedure works well when dealing with exact solutions
to the Einstein equations, in the sense that the joined solution is
itself also a solution. As we show in this paper, however, when
working with approximate solutions better transition functions need to
be found.

The purpose of this paper is to study whether a global approximate
solution to the Einstein equations, be it pure or mixed, can be
constructed directly from a weighted linear superposition of
approximate solutions with transition functions. We thus refine, prove
and verify many of the broad statements made
in~\cite{Yunes:2005nn,Yunes:2006iw} regarding transition functions.
This goal is achieved by constraining the family of allowed transition
functions for pure joined solutions via certain sufficient asymptotic
differentiability conditions. These conditions are independent of the
perturbative order to which asymptotic matching is carried out and the
location in which the transition occurs, as long as it is inside the
buffer zone. We then derive and prove theorems that guarantee that
pure joined solutions constructed with this restricted family of
transition functions satisfy the Einstein equations to the same order
as the approximations. Moreover, we also show that if the transition
functions do not satisfy the proposed conditions, their derivatives
modify the energy-matter content of the spacetime by introducing a
non-negligible stress-energy tensor.  Finally, we extend these
theorems to solutions projected onto spatial hypersurfaces, so that
they can be directly applied to initial data construction schemes.
These theorems then allow for the systematic construction of pure
joined solutions and they can be straightforwardly extended to mixed
joined solutions.

An example of the proposed theorems and allowed transition functions
is then provided by studying a binary system of non-spinning black
holes, where the approximate solutions are taken to be given by a
post-Newtonian expansion and a linearly perturbed Schwarzschild
solution. We shall not perform a systematic study of transition
functions here, but instead we pick functions that are variations of
those chosen in~\cite{Yunes:2005nn,Yunes:2006iw} in order to
illustrate how the gluing procedure works and how it breaks down. We
explicitly show that if the transition functions satisfy the proposed
conditions, the $4$-Ricci scalar calculated with the pure joined
solution vanishes to the same order as the uncontrolled remainders in
the approximations.  In particular, we explicitly show that
derivatives of the pure joined approximate solution built with
appropriate transition functions are equal to derivatives of both
original approximate solutions up to the uncontrolled remainders in
the approximations. We also numerically show that if the transition
functions violate the proposed conditions, their derivatives modify
the matter content of the spacetime by introducing a shell of matter.
In this manner, we explicitly verify, both analytically and
numerically, that a pure joined solution does represent the same
spacetime as that of the original approximate solutions in their
respective regions of validity up to the accuracy of the
approximations used.

This paper is divided as follows: in \sref{stan-junc-conds} we review
the standard junction conditions at a hypersurface, so that we can
extend them to the case where the solutions used are approximate
instead of exact, provided the existence of a buffer zone; in
\sref{asy-junc-conds} we study how to join approximate solutions with
transition function and derive conditions such that a pure joined
solution satisfies the Einstein equations to the same order as the
approximations used; in this section, we also study projections of
these pure joined solutions to a Cauchy hypersurface in order to
develop conditions for transition functions that can be used in
initial data construction schemes; in \sref{calc-trans-func} we study
an example of the theorems formulated by considering a binary system
of non-spinning black hole, constructing approximate global metrics
with different transition function, and explicitly calculating the
$4$-Ricci scalar; in \sref{conclusion} we conclude and point to future
research.

The notation of this paper is as follows: Greek indices range over all
spacetime indices, while Latin indices range only over spatial
indices; the symbol ${\cal{O}}(A)$ stands for terms of order $A$ at
most, while the symbol ${\cal{O}}(A,B)$ stands for remainders of order
$A$ or $B$ at most, where $A$ and $B$ are dimensionless; a tilde
superscript $\widetilde{A}$ stands for the asymptotic expansion of $A$
as defined in~\cite{Bender,Kevorkian}; the relation $A=B +
{\cal{O}}(C)$ means that $A$ is asymptotic to $B$ with uncontrolled
remainders of order $C$ (the so-called Landau or asymptotic notation);
we use units where $c = G = 1$. Symbolic calculations are performed
with either Mathematica or Maple.

\section{Junction Conditions at a Hypersurface}
\label{stan-junc-conds}

In this section, we review a variation of the standard junction
conditions of general relativity. These conditions have been discussed
extensively in the literature (see
\cite{Lanczos:1922,Lanczos:1924,Darmois:1927,Lichnerowicz,Misner:1964,Barrabes:1991ng,Israel:1966nc,MTW,poisson}
and references therein.) Here, we review only those concepts important
to the understanding of this paper, following in
particular~\cite{poisson}. Certain departures from the notation
of~\cite{poisson} are so that the generalization to the case of
approximate solutions in the next section becomes easier.

Let us first set up the problem. Consider a spatial (or timelike)
hypersurface $\cal{J}$ that divides spacetime into two regions:
${\cal{C}}_1$ and ${\cal{C}}_2$. Each of these regions possesses an
associated metric and coordinates,
$\{g_{\mu\nu}^{(1)},x_{(1)}^{\alpha}\}$ and $\{g_{\mu
  \nu}^{(2)},x_{(2)}^{\alpha} \}$, such that these metrics solve the
Einstein equations exactly in their respective regions. In the
literature, this exact solutions and coordinate systems are also
sometimes referred to as $\{g_{\mu \mu}^{\pm},x_{\pm}^{\alpha}\}$.
Let us further assume that an overlapping coordinate system
$x^{\alpha}$ exists in a neighbourhood of ${\cal{J}}$. The problem is
to formulate junction conditions on ${\cal{J}}$ that guarantee that
the joined $4$-metric satisfies the Einstein equations.

Let us make these statements more precise by considering a congruence
of geodesics piercing ${\cal{J}}$ defined with respect to first metric
in ${\cal{C}}_1$ and the second metric in ${\cal{C}}_2$ (see
\cite{poisson} for a detailed definition of such congruence.) Let then
$\ell$ denote the proper time (or proper distance) along the
geodesics, such that $\ell=0$ corresponds to when the geodesics reach
${\cal{J}}$. The joined solution then takes the following
form~\cite{poisson}:
\be
\label{joined-g}
g_{\mu \nu} = \Theta(\ell) g_{\mu \nu}^{(1)} + \Theta(-\ell)
g_{\mu\nu}^{(2)},
\ee
where $\Theta(\cdot)$ is the Heaviside function. Equation
(\ref{joined-g}) implicitly uses coordinates $x^{\alpha}$ that overlap
both the coordinates local to ${\cal{C}}_1$ and ${\cal{C}}_2$ in a
neighbourhood of ${\cal{J}}$.

We now proceed to formulate the junction conditions. The first
junction condition arises by requiring that the $4$-metric be
continuous across ${\cal{J}}$, in a coordinate system $x^{\alpha}$
that overlaps both $x^{\alpha}_1$ and $x^{\alpha}_2$ in an open region
that contains this hypersurface. This condition can be expressed in a
coordinate-independent way by projecting it to ${\cal{J}}$. Then, in
terms of $3$-tensors this condition becomes
\be
\label{first-junc-2}
\left.h_{ab}^{(1)}\right|_{{\cal{J}}} = 
\left.h_{ab}^{(2)}\right|_{{\cal{J}}},
\ee
where here $h_{ab}$ is the $3$-metric associated with the junction
hypersurface, {\textit{i.e.}}
\be
h_{\mu \nu} = g_{\mu \nu} + e_{\pm} n_{\mu} n_{\nu},
\ee
with $n_a$ normal to ${\cal{J}}$ and $e_{\pm} = \pm 1$ depending on
whether ${\cal{J}}$ is spatial ($-$) or timelike ($+$.)
\Eref{first-junc-2} guarantees that the hypersurface ${\cal{J}}$
has a well-defined geometry. These equations also imply that the
metric is differentiable across ${\cal{J}}$, except for its normal
derivative to ${\cal{J}}$ that in general is discontinuous.

The second junction condition arises by requiring that this normal
derivative does not lead to violations of the Einstein equations
across ${\cal{J}}$. In terms of $3$-tensors on ${\cal{J}}$, this
condition becomes
\be
\label{second-junc-2}
\left.K_{ab}^{(1)}\right|_{{\cal{J}}} = 
\left.K_{ab}^{(2)}\right|_{{\cal{J}}}.
\ee
One can show that the failure of these equations to be satisfied
changes the distribution of energy-momentum tensor of the spacetime
and gives raise to a shell of matter with stress-energy
tensor~\cite{poisson}
\be
\label{Tab}
S_{ab} = - \frac{\epsilon}{8 \pi} 
\left[  \left. K_{ab}^{(1)} \right|_{\cal{J}} - \left. K_{ab}^{(2)}
    \right|_{\cal{J}} - h_{ab} \left(  \left. K^{(1)}
      \right|_{\cal{J}} - \left. K^{(2)} \right|_{\cal{J}} \right)
  \right]. 
\ee
In the next section, we shall be mostly interested in vacuum
spacetimes, for which such a stress-energy tensor should vanish.

The satisfaction of the Einstein equations out of ${\cal{J}}$ by
\eref{joined-g} and the absence of a stress-energy tensor as in
\eref{Tab} then guarantees that the junction conditions
[\eref{first-junc-2} and \eref{second-junc-2}] are also satisfied.
Exact solutions, however, are rarely available for astrophysically
realistic scenarios. In that case, one must rely on approximate
solutions, for which similar conditions to those discussed here can be
found, as we shall study in the next section.

\section{Pure joined solutions}
\label{asy-junc-conds}
In this section we build a pure joined solution by extending the
standard junction conditions to the case where the metrics $g_{\mu
  \nu}^{(1,2)}$ are only approximate solutions to the Einstein
equations. For simplicity, we assume a vacuum spacetime and that there
exists analytic approximate expressions for $g_{\mu \nu}^{(1,2)}$,
such that pure joined solutions are sought. The conclusions of this
section, however, can straightforwardly be extended to other cases,
where numerical solutions are available instead of analytical ones,
provided information about first and second derivatives of the
numerical solutions is also available (note that for numerical
solutions the continuum derivative operator must be replaced by its
finite counterpart.) As shown in~\cite{Yunes:2005nn,Yunes:2006iw}, the
first step in joining approximate solutions is to apply asymptotic
matching inside some overlap region.  Once this has been done, one can
search for conditions such that the pure joined metric tensor
satisfies the Einstein equations to the same order as
$g_{\mu\nu}^{(1,2)}$. We here first present the basics of asymptotic
matching as applicable to this
problem~\cite{Blanchet:2002av,Yunes:2005nn}. We then search for
asymptotic junction conditions (or buffer zone conditions), which are
asymptotic in the sense of~\cite{Bender,Kevorkian} and, thus, are to
be understood only approximately to within some uncontrolled
remainder.

\subsection{Asymptotically matched metrics}


Consider then a manifold ${\cal{M}}$ that can be divided into two
submanifolds with boundary $C_1$ and $C_2$, each equipped with a
approximate metrics $g_{\mu \nu}^{(1)}$ and
$g_{\bar{\mu}\bar{\nu}}^{(2)}$ and a coordinate system $x^{\alpha}$
and $x^{\bar\alpha}$ respectively.  These metrics are approximate in
the sense that they solve the Einstein equations to
${\cal{O}}(\epsilon_{n}^{\ell_n})$ for some $l \in \mathbb{N}$ in
their respective submanifolds, {\textit{e.g.}} in vacuum
\be
G_{\mu \nu} [g_{\mu \nu}^{(n)}] = {\cal{O}}(\epsilon_{n}^{\ell_n+1}),
\ee
where $n = \{1,2\}$, $(\ell_1,\ell_2)$ are real numbers greater than
zero, and $\epsilon_n \ll 1$ is some dimensionless combination of
parameters and coordinates relative to the $n$-th submanifold. The
symbol ${\cal{O}}(\epsilon_{n}^{\ell_n + 1})$ refers to terms of
{\emph{relative}} order $\epsilon_n^{\ell_n+1}$ with respect to the
leading order term in the approximate solution $g_{\mu \nu}^{(n)}$.
In principle, there could be logarithms of $\epsilon_n$ present in the
remainders, such as in high-order post-Newtonian expansions, but we
neglect such terms here because they shall not affect the analysis of
this paper. Notice that we use here bars to denote the different
coordinate systems (as opposed to numbers, as in the previous section)
because we must be more careful about the coordinates used in these
approximate metric components.  For concreteness, let the region of
validity of $g_{\mu\nu}^{(1)}(x^{\alpha})$ be defined by $x^{\alpha}
\gg x^{\alpha}_{in}$ and that of $g_{\mu \nu}^{(2)}(x^{\bar\alpha})$
by $x^{\bar\alpha} \ll x^{\bar\alpha}_{out}$. These inequalities
define a {\textit{spacetime}} region of validity, since the
approximate metric might not be valid for all times. For example, such
is the case for the post-Newtonian metric of two point particles in
quasi-circular orbit, which is valid only for times $t \ll t_c$, where
$t_c$ is the time of coalescence.

Let us further assume that these submanifolds overlap in some
$4$-volume, defined by the intersection ${\cal{B}} = C_1 \cap C_2$,
and sometimes referred to as the overlap region or buffer zone. The
boundary of the buffer zone, $\partial{\cal{B}}$, cannot be determined
exactly, because it is inherently tied to the regions of validity of
the approximate solutions, which themselves are only defined
approximately. With this in mind, let us further assume that the
charts $\{x^{\alpha}\}$ and $\{x^{\bar{\alpha}}\}$ are defined in the
neighbourhood of any field point in the buffer zone and that they
satisfy $x^{\bar\alpha} = x^{\alpha} +
{\cal{O}}(\epsilon_1^{\ell_1+1}, \epsilon_2^{\ell_2+1})$.  The buffer
zone can then be asymptotically defined by the following condition:
$(x^{\alpha}_{in} x_{\alpha}^{in})^{1/2} \ll (x^{\alpha}
x_{\alpha})^{1/2} \ll (x^{\alpha}_{out} x_{\alpha}^{out})^{1/2}$,
where indeces are raised or lowered with the local metric to
${\cal{C}}_{1,2}$. Wherever possible, we use the Landau notation,
which specifically specifies the behavior of the remainder.  The
definition of the boundary of the buffer zone should be understood
only in an asymptotic sense, as defined in~\cite{Bender,Kevorkian}.
This boundary is made up of two disconnected pieces,
$\partial{\cal{B}}_-$ and $\partial{\cal{B}}_+$, defined via
$x^{\alpha} = x_{in}^{\alpha} + {\cal{O}}(\epsilon_1^{\ell_1+1},
\epsilon_2^{\ell_2+1})$ and $x^{\alpha} = x_{out}^{\alpha} +
{\cal{O}}(\epsilon_1^{\ell_1+1}, \epsilon_2^{\ell_2+1})$ respectively.
The definition of the buffer zone can be thought of in terms of a
simplistic spherically symmetric static example, where $x^{\alpha} \to
r$, $\partial{\cal{B}}$ is a spherical shell and
$\partial{\cal{B}}_{\pm}$ are $2$-spheres.  However, in practical
examples, such as binary black hole spacetimes, the boundary of the
buffer zone is not simply a spherical shell, but instead it acquires
some deformation in accordance with the deformation of the spacetime
that is being modeled.

Before proceeding with the description of asymptotic matching, let us
make some comments on the approach adopted in this paper. In the
previous paragraphs and in what shall follow, we have not adopted a
rigorous geometrical approach in the description of asymptotic
matching, the definition of the buffer zone and the submanifolds. A
rigorous geometrical approach, for example, would impose other
constraints on ${\cal{B}}$, such that it is properly embedded in
${\cal{C}}_1$ and ${\cal{C}}_2$.  Furthermore, such an approach would
describe the conditions under which asymptotic matching can be
performed, since in general this method is valid only locally.
Instead of such an approach, in this paper we adopt an
{\emph{analytical}} one, where we only provide a minimal amount of
details in order to make the presentation simpler.  For example, we
shall restrict our attention to coordinate grids that coincide when
both $\epsilon_1 \to 0$ and $\epsilon_2 \to 0$ and, henceforth, we
shall call ${\cal{C}}_{1,2}$ regions instead of submanifolds. Such an
approach is adopted because the aim of this paper is not to provide a
detailed geometrical account of asymptotic matching, but instead to
study transition functions and approximate joined solutions (see
\cite{Poujade:2001ie} and references therein for a more detailed and
rigorous geometric account of asymptotic matching.)

The approximate metrics live in different regions and depend on
different coordinates ($x^{\alpha}$,$x^{\bar{\alpha}}$) and local
parameters ($\theta^{\alpha}$,$\theta^{\bar{\alpha}}$.) Examples of
these parameters are the mass of the system and its velocity. However,
since both $g_{\mu \nu}^{(n)}$ are valid in ${\cal{B}}$, both
coordinate systems must be valid in the buffer zone ({\textit{i.e.}},
the charts of $C_n$ overlap in ${\cal{B}}$.)  Using the uniqueness
theorems of asymptotic expansions~\cite{Bender,Kevorkian}, one can
find a coordinate and parameter transformation to relate adjacent
regions inside ${\cal{B}}$. In order to achieve this, one must first
compute the asymptotic expansions of the approximate line elements,
$\widetilde{ds}^2_{(n)}$, near the boundaries of the buffer zone,
$\partial{\cal{B}}$, and then compare them inside ${\cal{B}}$ but yet
away from $\partial {\cal{B}}$, {\textit{i.e.}}
\be
\label{asympt-cond}
\widetilde{ds}^2_{(1)} -  \widetilde{ds}^2_{(2)} =
{\cal{O}}(\epsilon_1^{\ell_1+1}, \epsilon_2^{\ell_2+1}), \qquad
\textrm{\mbox{in}} \quad {\cal{B}} \; \backslash \; {\partial{\cal{B}}},  
\ee
where the $\backslash$ symbol is the standard exclusion symbol of set
theory and where ${\cal{O}}(\epsilon_1^{\ell_1+1},
\epsilon_2^{\ell_2+1})$ stands for uncontrolled remainders of order
$\epsilon_1^{\ell_1+1}$ {\emph{or}} $\epsilon_2^{\ell_2+1}$ as defined
at the end of Sec.~1. Using \eref{asympt-cond}, a coordinate
transformation $\psi: x^{\bar\alpha} \to x^{\alpha}$ and a parameter
transformation $\phi: \theta^{\bar\alpha} \to \theta^{\alpha}$ can be
found inside ${\cal{B}}$. We have written~(\ref{asympt-cond}) in terms
of the line element, but we could have easily written it in terms of
the metric components. In fact, depending on the calculation, it might
be necessary to asymptotically match different components of the
metric to different order. For example, for the construction of
initial data, the $g_{00}$ and $g_{ij}$ components of the metric need
to be matched to lower order than the $g_{0i}$
components~\cite{Yunes:2005nn,Yunes:2006iw}. In what follows, we will
assume that all metric components have been asymptotically matched to
the same order. When this is not the case, the conditions that we
shall propose on transition functions can be adapted by noting that
one must use the highest order to which metric components have been
asymptotically matched.

These transformations guarantee that all components of adjacent
$4$-metrics are asymptotic to each other inside the buffer
zone~\cite{Bender,Kevorkian}. This fact implies that the derivatives
of adjacent metrics are also asymptotic to each other
\be
\label{asympt-cond-deriv}
\frac{\partial^m}{\partial x^{\alpha_1} \ldots \partial x^{\alpha_m}}
g_{\mu \nu}^{(1)} - \frac{\partial^m}{\partial x^{\alpha_1} \ldots
  \partial x^{\alpha_m}}  g_{\mu \nu}^{(2)} =
{\cal{O}}(\epsilon_1^{\ell_1+1-m}, \epsilon_2^{\ell_2+1-m}), 
\ee 
for all $m < l$. However, the relative order to which the derivatives
of adjacent metrics are asymptotic to each other is in general not the
same as the relative order to which the adjacent metrics themselves
are asymptotic to each other. This reduction in {\emph{matching
    accuracy}} is due to the implicit assumption that asymptotic
matching is carried out to some finite order.
In~(\ref{asympt-cond-deriv}), we have assumed that this decrease in
accuracy occurs in single powers of $\epsilon_{1,2}$, but if this is
not the case the results of this paper can be rescaled appropriately.

Let us provide a general example of such matching accuracy reduction,
without specifying a particular spacetime (for a more detailed example
see~\cite{Yunes:2005nn,Yunes:2006iw}.) Consider a post-Newtonian
expression that is known to ${\cal{O}}(1/c^4)$ and that is
asymptotically matched to a perturbed black hole solution in some
buffer zone to ${\cal{O}}(1/c^4,\epsilon^2)$, where $c$ is here the
speed of light and $\epsilon$ is the black hole perturbation theory
expansion parameter. Let us now take the time derivative of the
post-Newtonian expression, which results in a term of
${\cal{O}}(1/c^5)$. In post-Newtonian theory, one would never truncate
the differentiated term at ${\cal{O}}(1/c^4)$. However, in the theory
of asymptotic matching to a finite order, the time derivative of the
post-Newtonian quantity remains matched only to ${\cal{O}}(1/c^4,
\epsilon^2)$, and thus it will disagree with the perturbed black hole
quantity at ${\cal{O}}(1/c^5)$.  It is in this sense that
differentiation usually decreases the relative order to which two
expression have been asymptotically matched (we should note that when
asymptotic matching is carried out to all orders, then this reduction
in matching accuracy disappears~\cite{Poujade:2001ie}.)

Let us provide a more explicit example that, although similar in
spirit to the one described above, does not require general
relativity. Consider two functions 
\be
f(t) = \left[f_0^2 + 2 f_1^2 \epsilon_1 x(\omega \; t)\right]^{1/2},
\qquad 
g(t) = \frac{9}{9 + \epsilon_2 y(\Omega \; t)},
\ee
where $f_{0,1}$ are constants and $x(\omega t)$ and $y(\Omega t)$ are
periodic funtions of time with period $1/\omega$ and $1/\Omega$
respectively. Let us pretend that $f(t)$ and $g(t)$ are approximate
solutions to the same differential equation in the limits
$\epsilon_{1} \ll 1$ and $\epsilon_{2} \ll 1$ respectively. If we
asymptotically match these functions in a buffer zone where both
$\epsilon_1 \ll 1$ and $\epsilon_2 \ll 1$ up to uncontrolled
remainders of ${\cal{O}}(\epsilon_1^2,\epsilon_2^2)$, we discover that
at $t=0$ the constants are $f_0=1$ and $f_1 = y(0)/[9 x(0)]$. If we
now study their time derivatives, we find that at $t=0$
\be
\frac{\partial f}{\partial t} = \epsilon_1 \frac{y(0)}{9 x(0)}
\omega \left(\frac{\partial x}{\partial t}\right)_{t=0},
\qquad 
\frac{\partial g}{\partial t} = \frac{\epsilon_2}{9} \Omega
\left(\frac{\partial y}{\partial t}\right)_{t=0}, 
\ee
with uncontrolled remainders of
${\cal{O}}(\epsilon_1^2,\epsilon_2^2)$. Note that the derivatives are
not equal to each other to this order because the velocities $\omega$
and $\Omega$ have not yet been determined via asymptotic matching.
The reduction in matching accuracy is here explicit since, while
$f(t)$ and $g(t)$ have been asymptotically matched up to uncontrolled
remainders of ${\cal{O}}(\epsilon_1^2,\epsilon_2^2)$, their time
derivatives match only up to uncontrolled remainders of
${\cal{O}}(\epsilon_1,\epsilon_2)$.

Henceforth, we assume that asymptotic matching has been carried out
and that one of the metrics has been transformed according to
$\{\psi,\phi\}$ such that \eref{asympt-cond} holds.  We refer the
reader to~\cite{Bender,Kevorkian} for more details on the theory of
asymptotic analysis
and~\cite{Blanchet:2002av,Yunes:2005nn,Burke,Burke-Thorne,Death:1974o,Death:1975,Thorne:1984mz}
for a more detailed discussion on asymptotic expansions in general
relativity.

\subsection{Asymptotic Junction Conditions}
Let us now return to the buffer zone and note that it can be foliated
by a family of junction hypersurfaces ${\cal{J}}_i$, where $i$ labels
the member of the family. For simplicity, we choose these
hypersurfaces to be timelike, since in the next section we project the
$4$-metric to a Cauchy (spatial) hypersurface $\Sigma$ and it is
convenient then that ${\cal{J}}_i$ be orthogonal to $\Sigma$.
Asymptotic matching has provided a coordinate transformation to relate
the charts inside ${\cal{B}}$, so, in particular, these matched
coordinates are valid in an open region containing every ${\cal{J}}_i$
as long as this is not close to ${\partial{\cal{B}}}$.  We look for
asymptotic junction conditions in this subregion of ${\cal{B}}$
({\textit{i.e.}}  away from ${\partial{\cal{B}}}$), in terms of
differentiability conditions of $g_{\mu\nu}^{(n)}$.  

Furthermore, let us also consider a family of geodesic congruences
$\gamma_i$ that pierce ${\cal{J}}_i$. The $i$th member of the family
is parameterized by proper distance $\ell_i$ to the $i$th hypersurface
${\cal{J}}_i$, such that $\ell_i = 0$ occurs when that member reaches
that hypersurface. Such a family of geodesic congruences is defined
with respect to the approximate metrics $g_{\mu \nu}^{(n)}$ since
these metrics are equal to each other up to uncontrolled remainders in
the asymptotic matching scheme. Even though these geodesics are not
strictly necessary for the construction of asymptotic junction
conditions, we find them useful to define a measure of distance to
${\cal{J}}_i$, which shall later become important in the definition of
transition functions.

We can now begin to look for asymptotic junction conditions by
considering the following $4$-metric tensor:
\ba
\label{joined-g-asy}
g_{\mu \nu} = F(\ell_i) \; g_{\mu \nu}^{(1)} + \left[1 -
  F(\ell_i)\right] \;  g_{\mu \nu}^{(2)}, 
\ea
where $F(\cdot)$ is a (proper) transition function that smoothly
ranges from zero to unity inside ${\cal{B}}$ and which will be defined
more rigorously later.  \Eref{joined-g-asy} is motivated by the fact
that any tensor $E^{a_1 a_2 \ldots a_i}{}_{b_1 b_2 \ldots b_j}$ can
always be split into
\ba
\label{split}
E^{a_1 a_2 \ldots a_i}{}_{b_1 b_2 \ldots b_j} &=& G^{a_1 a_2 \ldots
  a_i}{}_{b_1 b_2 \ldots b_j} F(\ell_i) 
\nonumber \\
&+&  \left[1 - F(\ell_i)\right] H^{a_1
  a_2 \ldots a_i}{}_{b_1 b_2 \ldots b_j},  
\ea
provided that
\be
E^{a_1 a_2 \ldots a_i}{}_{b_1 b_2 \ldots b_j} = G^{a_1 a_2
  \ldots a_i}{}_{b_1 b_2 \ldots b_j} = H^{a_1 a_2 \ldots a_i}{}_{b_1
  b_2 \ldots b_j}
\ee
inside the transition region and that the proper transition function
is sufficiently regular. The regularity requirement is to guarantee
that both terms of \eref{split} are differentiable.  In
\eref{joined-g-asy} this split is valid because $g_{\mu \nu}^{(1)}$ is
asymptotic to $g_{\mu\nu}^{(2)}$ in ${\cal{B}}$ up to uncontrolled
remainders once the maps $\{\psi,\phi\}$ have been applied. As we will
see below, however, care must be taken when constructing such proper
transition functions $F(\cdot)$ to avoid ruining the differentiability
properties of the joined metric.

Let us now define proper transition functions in order to clarify how
to merge the metrics via \eref{joined-g-asy}. A proper transition
function is a smooth real map ${\cal{F}}: \Re \to \left[0,1\right]$
that ranges from zero to unity inside some transition window $w$,
while it acquires the value of $1/2$ as $\ell_i \to 0$, and that
satisfies the following conditions:
\ba
\label{asy-trans}
F(\ell_i) &\to& 1 \quad {\textrm{as}} \quad \ell_i \to \ell_{+},
\nonumber \\
F(\ell_i) &\to& 0 \quad {\textrm{as}} \quad \ell_i \to \ell_{-}, 
\ea
where here $\ell_{\pm}$ is the proper distance to
$\partial{\cal{B}}_{\pm}$. Note that the point $\ell_i = 0$ is where
the global approximate metric of \eref{joined-g-asy} contains equal
contributions from $g_{\mu \nu}^{(1)}$ and $g_{\mu \nu}^{(2)}$. The
transition window will be studied later in Sec.~4, but we can think of
it qualitatively as a parameter of a proper transition function that
determines the region where these functions are significantly
different from unity or zero. The requirement that a proper transition
function vanishes or tends to unity at $\partial{\cal{B}}_{\pm}$ is
necessary to avoid contamination of $g_{\mu\nu}^{(1)}$ in
${\cal{C}}_2$ and vice versa, since in general $g_{\mu\nu}^{(n)}$ has
large uncontrolled remainders and could diverge outside its region of
validity. In fact, the speed at which a proper transition function
must tend to unity or zero will depend on the speed of the growth of
the uncontrolled remainders of the approximations outside their region
of validity, as we shall study later. This definition does not
constrain how the transition function behaves inside the transition
region. Also note that this family of functions tends to the Heaviside
function of the previous section as $w \to 0$ and that it does not
need to be analytic.  We shall not constrain this family further for
now, but instead we search for conditions on this family such that
\eref{joined-g-asy} satisfies the Einstein equations to the same order
as $g_{\mu\nu}^{(n)}$.

With these proper transition functions, we immediately see that the
joined metric itself also satisfies the following set of asymptotic
conditions:
\ba
g_{\mu \nu} &=& g_{\mu \nu}^{(1)} + {\cal{O}}(\epsilon_1^{\ell_1+1},
\epsilon_2^{\ell_2+1}) , \quad  {\textrm{in}} \quad 
{\cal{C}}_1 \ \backslash \ {\cal{B}}, 
\nonumber \\
g_{\mu \nu} &=& g_{\mu \nu}^{(2)} +
{\cal{O}}(\epsilon_1^{\ell_1+1}, \epsilon_2^{\ell_2+1}), \quad
{\textrm{in}} \quad {\cal{C}}_2 \ \backslash \ {\cal{B}}, 
\ea
while in the buffer zone the metric is some weighted linear
superposition of both approximate solutions. Thanks to asymptotic
matching, the approximate metrics are identical inside the buffer zone
up to uncontrolled remainders [see \eref{asympt-cond-deriv}] and,
thus, this linear superposition is valid there, in spite of the
non-linearity of the Einstein field equations. Also note that we are
free to choose any junction hypersurface ${\cal{J}}_i$ to join the
metrics as long as it is inside ${\cal{B}}$ but away from
$\partial{\cal{B}}$. However, there is usually a typical choice of
${\cal{J}}_i$, given by the surface where the error bars of the
approximate metrics become comparable. Such a choice is not unique,
but has previously proved to be close to optimal in certain
scenarios~\cite{Yunes:2005nn,Yunes:2006iw}.  Asymptotic matching then
seems to be a good technique for the construction of a pure joined
$4$-metric as given in \eref{joined-g-asy}, as long as we find
transition functions that are sufficiently well-behaved so that their
derivatives do not introduce errors larger than those contained in the
approximations.

We now proceed to determine the asymptotic junction conditions by
analogy with the standard junction conditions discussed in the
previous section. The first junction condition of the previous section
is automatically satisfied asymptotically in ${\cal{B}}$ via
\eref{asympt-cond} with uncontrolled remainders of
${\cal{O}}(\epsilon_1^{\ell_1+1},\epsilon_2^{\ell_2+1})$.  However,
the second junction condition is not necessarily satisfied because
\eref{asympt-cond} does not guarantee differentiability across
${\cal{J}}_i$ to the same order as continuity. Differentiating
\eref{joined-g-asy} we obtain
\be
\label{first-deriv}
g_{\mu \nu, \alpha} = F(\ell_i) \; g_{\mu \nu,\alpha}^{(1)} + \left[ 1 -
  F(\ell_i) \right] \; g_{\mu \nu,\alpha}^{(2)} + F_{,\alpha}
  \left\{g_{\mu\nu}\right\}, 
\ee
where we have defined the operation $\left\{E\right\} \equiv E^{(1)} -
E^{(2)}$ for any function $E$. In order for \eref{joined-g-asy} to be
a solution to the Einstein equations, we should require that the third
term be as small as the uncontrolled remainders of the first two
terms.  By \eref{asympt-cond-deriv}, we know that the last piece of
the third term is of
${\cal{O}}(\epsilon_1^{\ell_1+1},\epsilon_2^{\ell_2+1})$.  On the
other hand, the first two terms are bounded below by their smallest
size, which is of
${\cal{O}}(\epsilon_1^{\ell_1},\epsilon_2^{\ell_2})$. We, thus, arrive
at the condition
\be
\label{first-deriv-cond}
F_{,\alpha} = {\cal{O}}(\epsilon_1^{0},\epsilon_2^{0}), \quad
{\textrm{in}} \; {\cal{B}}. 
\ee
One can show that this condition is sufficient, since it excludes
cases where $F_{,\alpha} = {\cal{O}}(\ln(\epsilon_1,\epsilon_2))$.
With this condition, \eref{first-deriv} becomes
\be
\label{first-deriv-asy}
g_{\mu \nu, \alpha} = F(\ell_i) \; g_{\mu \nu,\alpha}^{(1)} + \left[1
  - F(\ell_i)\right]  \; g_{\mu \nu,\alpha}^{(2)} +
{\cal{O}}(\epsilon_1^{\ell_1+1},\epsilon_2^{\ell_2+1}).  
\ee

In vacuum, another condition must be imposed on the $4$-metric in
order for the Einstein equations to be asymptotically satisfied inside
${\cal{B}}$. This condition can be enforced by requiring that the
$4$-metric tensor be asymptotically $C^2$ in ${\cal{B}}$.
Differentiating \eref{first-deriv} we obtain
\ba
\label{second-deriv}
g_{\mu \nu, \alpha \beta} &=& F(\ell_i) \; g_{\mu \nu,\alpha\beta}^{(1)}
+ \left[1 - F(\ell_i) \right] \; g_{\mu \nu,\alpha\beta}^{(2)}
\nonumber \\
&+& 
  2 F_{,(\alpha} \left\{g_{\mu\nu,|\beta)}\right\} + F_{,\alpha\beta}
 \left\{g_{\mu\nu}\right\}.  
\ea
Here the parenthesis on the indices represent the standard symmetry
operation $g_{(a|b,|c)} = 1/2 \left( g_{ab,c} + g_{cb,a} \right)$.  In
order for the metric to be asymptotically $C^2$ in ${\cal{B}}$, we
must require that the last two terms be much smaller than the first
two. This requirement can be enforced by requiring that
\ba
\label{sec-cond}
F_{,\alpha} &=& {\cal{O}}(\epsilon_1,\epsilon_2), \quad {\textrm{in}} \; {\cal{B}},
\nonumber \\
F_{,\alpha\beta} &=& {\cal{O}}(\epsilon_1^{0},\epsilon_2^{0}),
\quad {\textrm{in}} \; {\cal{B}}. 
\ea
The first condition in \eref{sec-cond} is a refinement of
\eref{first-deriv-cond}, while the second condition is new.  These
equations are also compatible with $F_{,\alpha} \ll F_{,\alpha\beta}$,
which is a consequence of the fact that $\left\{g_{\mu
    \nu,\alpha}\right\} =
{\cal{O}}(\epsilon_1^{\ell_1},\epsilon_2^{\ell_2})$. As one can show,
the decrease in matching accuracy of the derivatives of matched
expressions has lead to different conditions for the first and second
derivatives of the transition functions. In this way, the last two
terms of \eref{second-deriv} become of
${\cal{O}}(\epsilon_1^{\ell_1+1},\epsilon_2^{\ell_2+1})$ or smaller,
because by \eref{asympt-cond-deriv}, $\left\{g_{\mu \nu}\right\} =
{\cal{O}}(\epsilon_1^{\ell_1+1},\epsilon_2^{\ell_2+1})$ and
$\left\{g_{\mu \nu,\alpha}\right\} =
{\cal{O}}(\epsilon_1^{\ell_1},\epsilon_2^{\ell_2})$.  With these
conditions, \eref{second-deriv} becomes
\be
g_{\mu \nu, \alpha \beta} = F(\ell_i) \; g_{\mu \nu,\alpha\beta}^{(1)} 
+ \left[1 - F(\ell_i)\right]  \; g_{\mu \nu,\alpha\beta}^{(2)} +
{\cal{O}} (\epsilon_1^{\ell_1+1},\epsilon_2^{\ell_2+1}).  
\ee

Transition functions that satisfy \eref{sec-cond} are enough to
guarantee a $C^2$ global metric in ${\cal{B}}$, irrespective of the
matching order or the junction hypersurface chosen. Requiring the
joined solution to be $C^2$ in ${\cal{B}}$ guarantees that the
Einstein tensor is also continuous across any ${\cal{J}}_i$.  Since we
are dealing with a vacuum spacetime, the junction hypersurfaces do not
represent any physical boundary, such as a shell of matter, and can
thus be chosen arbitrarily as long as they are inside ${\cal{B}}$ and
away from $\partial{\cal{B}}$, so that the approximations are still
valid.

We have then derived conditions on the $4$-metric and, thus, on the
transition functions, that guarantee that the Einstein tensor is
continuous in ${\cal{B}}$. The standard junction conditions in the
presence of matter do not necessarily require the metric to be $C^1$
across ${\cal{J}}$, whereas here we must require it to be at least
asymptotically $C^2$ in ${\cal{B}}$, so that no artificial features
are introduced at the boundary. We can now formulate the following
theorem:
\begin{theorem}[Buffer Zone Junction Condition]
  Consider a spacetime manifold ${\cal{M}}$ that can be divided into
  two submanifolds with boundary $C_1$ and $C_2$, inside which
  $g^{(1,2)}_{\mu \nu}$ are approximate solutions to the vacuum
  Einstein equations up to uncontrolled remainders of
  ${\cal{O}}(\epsilon_1^{\ell_1+1},\epsilon_2^{\ell_2+1})$. Let the
  boundaries of the submanifolds be defined asymptotically by the
  approximate boundary of the region of validity of $g_{\mu
    \nu}^{(1,2)}$ and let the intersection of these submanifolds,
  ${\cal{B}} = C_1 \cap C_2$, be called the buffer zone. Let us
  foliate the buffer zone with a family of timelike hypersurfaces
  ${\cal{J}}_i$ and consider a family of geodesic congruences
  $\gamma_i$, such that the $i$th member is parametrized by proper
  distance $\ell_i$ to the $i$th hypersurface ${\cal{J}}_i$. Consider
  the family of joined $4$-metric tensor parameterized by $\ell_i$,
  namely
\be
\label{joined-g-asy-thm}
g_{\mu \nu} = F(\ell_i) \; g_{\mu \nu}^{(1)} + \left[1 - F(\ell_i)\right] \;
g_{\mu\nu}^{(2)},
\ee
where $F(\cdot)$ is a proper transition function as defined by
\eref{asy-trans}. Then, equation (\ref{joined-g-asy-thm}) is also an
$\ell_i$-independent approximate solution to the Einstein equations up
to uncontrolled remainders of
${\cal{O}}(\epsilon_1^{\ell_1+1},\epsilon_2^{\ell_2+1})$ if and only
if the following conditions hold:
\ba
&(i)& \qquad {\textrm{The metrics $g_{\mu\nu}^{(1,2)}$ have been
    asymptotically}}
\nonumber \\
&& \qquad \; {\textrm{matched in ${\cal{B}}$ up to uncontrolled
    remainders }}
\nonumber \\
&& \qquad \; {\textrm{of
    ${\cal{O}}(\epsilon_1^{\ell_1+1},\epsilon_2^{\ell_2+1})$}},  
\nonumber \\
&(ii)& \qquad F_{,\alpha} = {\cal{O}}(\epsilon_1,\epsilon_2), \quad
\textrm{in} \quad {\cal{B}},  
\nonumber \\ \nonumber 
&(iii)& \qquad F_{,\alpha\beta} = {\cal{O}}(\epsilon_1^{0},\epsilon_2^{0}),
\quad \textrm{in} \quad {\cal{B}}. 
\ea
\end{theorem}

\begin{proof}
  The proof follows directly from the calculation of the Ricci tensor,
  which should vanish in vacuum. Let us then use the joined metric to
  calculate the Christoffel connection. Doing so we obtain
\ba
\label{con}
\Gamma^{\alpha}_{\beta \gamma} &=& \Gamma^{\alpha \; (1)}_{\beta \gamma}
F(\ell_i) + \Gamma^{\alpha \; (2)}_{\beta \gamma} \left[1 - F(\ell_i)\right]
\\ \nonumber 
&+& 
 \frac{1}{2} g^{\alpha\delta} \left( F_{,\gamma} \left\{g_{\beta \delta}\right\} +
  F_{,\beta} \left\{ g_{\gamma \delta}\right\} - F_{,\delta}
  \left\{g_{\beta \gamma}\right\} \right),
\ea
where the $\{\cdot\}$ operator was defined in \eref{first-deriv}. We
have here used the fact that the inverse metric can be written as
$g^{\alpha \beta} = F \; g^{\alpha \beta}_{(1)} + (1 - F) \; g^{\alpha
  \beta}_{(2)}$, neglecting terms that are proportional to $F (1 - F)$
because they are of ${\cal{O}}(\epsilon_1^{2 \ell_1},\epsilon_2^{2
  \ell_2})$.  Outside the buffer zone the terms in parenthesis clearly
vanish, but inside this region they could be large. However, note that
by condition $(i)$, $\left\{g_{\mu \nu}\right\} =
{\cal{O}}(\epsilon_1^{\ell_1+1}, \epsilon_2^{\ell_2+1})$, while by
condition $(ii)$ the derivative of the transition function satisfies
$F_{,\alpha} = {\cal{O}}(\epsilon_1,\epsilon_2)$. Thus, the term in
parenthesis satisfies
\be
\label{first-curly-brace}
 \left( F_{,\gamma} \left\{g_{\beta \delta}\right\} +
  F_{,\beta} \left\{ g_{\gamma \delta}\right\} - F_{,\delta}
  \left\{g_{\beta \gamma}\right\} \right) \ll
  {\cal{O}}(\epsilon_1^{\ell_1+1},\epsilon_2^{\ell_2+1}).
\ee
Due to the precision of the Christoffel symbols, namely that
$\Gamma^{\alpha \; (1)}_{\beta \gamma} = \Gamma^{\alpha \; (2)
}_{\beta \gamma}$ in ${\cal{B}}$ up to uncontrolled remainders of
${\cal{O}}(\epsilon_1^{\ell_1+1},\epsilon_2^{\ell_2+1})$, combined
with conditions $(i)$ and $(ii)$, we are allowed to write the
connection as
\be
\label{connection-asy}
\Gamma^{\alpha}_{\beta \gamma} = \Gamma^{\alpha \; (1)}_{\beta \gamma}
F(\ell_i) + \Gamma^{\alpha \; (2)}_{\beta \gamma} \left[1 - F(\ell_i)\right] +
{\cal{O}}(\epsilon_1^{\ell_1+1},\epsilon_2^{\ell_1+1}).
\ee

However, in order to compute the $4$-Riemann tensor, we need the
derivative of the connection. This quantity is given by
\ba
\label{deriv-con}
\Gamma^{\alpha}_{\beta \gamma,\sigma} &=& \Gamma^{\alpha \;
  (1)}_{\beta \gamma,\sigma} F(\ell_i) + \Gamma^{\alpha \; (2)}_{\beta
  \gamma,\sigma} \left[1 - F(\ell_i)\right]  + F_{,\sigma}
\left\{\Gamma^{\alpha}_{\beta \gamma}\right\}  
\\ \nonumber 
&+& 
 \frac{1}{2} g^{\alpha\delta} 
  \left( F_{,\gamma \sigma} \left\{g_{\beta \delta}\right\} +
  F_{,\sigma \beta} \left\{ g_{\gamma \delta}\right\} - F_{,\delta
    \sigma} \left\{g_{\beta \gamma} \right\} \right)
\\ \nonumber 
&+&  \frac{1}{2} g^{\alpha \delta} 
 \left(F_{,\gamma} \left\{g_{\beta \delta,\sigma}\right\} +
  F_{,\beta} \left\{ g_{\gamma \delta,\sigma}\right\} - F_{,\delta}
  \left\{g_{\beta \gamma,\sigma} \right\} \right),
\ea
where other terms either vanish by \eref{first-curly-brace} or are
negligible.  \Eref{deriv-con} must be obtained by differentiating
\eref{con}, rather than \eref{connection-asy}, because otherwise we
would miss the second derivatives of the transition functions. The
third and fourth terms of \eref{deriv-con} are already of
${\cal{O}}(\epsilon_1^{\ell_1+1},\epsilon_2^{\ell_2+1})$ by
\eref{asympt-cond-deriv}, condition $(i)$, $(ii)$ and $(iii)$. The
last term is finally also of
${\cal{O}}(\epsilon_1^{\ell_1+1},\epsilon_2^{\ell_2+1})$ because
although $\left\{g_{\mu\nu,\sigma}\right\} =
{\cal{O}}(\epsilon_1^{\ell_1},\epsilon_2^{\ell_2})$, by condition
$(ii)$ $F_{,\alpha} = {\cal{O}}(\epsilon_1,\epsilon_2)$. We are
therefore left with
\be
\label{deriv-connection-asy}
\Gamma^{\alpha}_{\beta \gamma,\sigma} = \Gamma^{\alpha \;
  (1)}_{\beta \gamma,\sigma} F(\ell_i) + \Gamma^{\alpha \; (2)}_{\beta
  \gamma,\sigma} \left[1 - F(\ell_i)\right]  +
{\cal{O}}(\epsilon_1^{\ell_1+1},\epsilon_2^{\ell_2+1}), 
\ee
which again leads to the asymptotic condition $\Gamma^{\alpha \;
  (1)}_{\beta \gamma,\sigma} - \Gamma^{\alpha \; (2)}_{\beta
  \gamma,\sigma} =
{\cal{O}}(\epsilon_1^{\ell_1+1},\epsilon_2^{\ell_2+1})$ in
${\cal{B}}$.

We are now ready to compute the Ricci tensor. Once more, using
\Eref{connection-asy} and \eref{deriv-connection-asy} we can write
\be
R_{\alpha \beta} = R^{(1)}_{\alpha \beta} \; F(\ell_i) +
R^{(2)}_{\alpha \beta} \; \left[1 - F(\ell_i)\right] +
{\cal{O}}(\epsilon_1^{\ell_1+1},\epsilon_2^{\ell_2+1}), 
\ee
where the third term groups all the cross terms that are
${\cal{O}}(\epsilon_1^{\ell_1+1},\epsilon_2^{\ell_2+1})$ in
${\cal{B}}$. Furthermore, we know that the approximate solutions
$g_{\mu\nu}^{(1,2)}$ satisfy the Einstein equations to
${\cal{O}}(\epsilon_n^{\ell_n})$, which then implies that
\be
R^{(n)}_{\alpha \beta} = {\cal{O}}(\epsilon_{n}^{\ell_n+1}).
\ee
We thus arrive at the conclusion that 
\be
R_{\alpha \beta} = {\cal{O}}(\epsilon_1^{\ell_1+1},\epsilon_2^{\ell_2+1}), 
\ee
which then proves the theorem.
\end{proof}

This theorem allows for the construction of pure joined solutions,
with a restricted class of transition functions. Although the theorem
has been formulated for vacuum spacetimes, it also holds for
non-vacuum scenarios as discussed above. The proof for the non-vacuum
case can be established simply by following the above proof and
realizing that now $R^{(n)}_{\alpha \beta} - 8 \pi T^{(n)}_{\alpha
  \beta} = {\cal{O}}(\epsilon_{n}^{\ell_n+1})$.  Also note that the
ideas of this section can be extended to mixed joined solutions, by
replacing condition $(i)$ by some other condition that guarantees that
the approximate solutions represent the same spacetime in ${\cal{B}}$.
Finally, note that the conditions we impose on the transition
functions are not very stringent, thus allowing for a wide range of
possible functions.

\subsection{Projection to a Cauchy Hypersurface}
\label{proj-metric}
Let us now specialize the pure joined solution of \eref{joined-g-asy}
to joined initial data on a Cauchy hypersurface. Since this
hypersurface is by definition spacelike, it is convenient to have a
foliation of the buffer zone that is timelike as given in the previous
section. The data constructed in this section consists of an induced
$3$-metric on the Cauchy hypersurface and its extrinsic curvature.

Consider then a Cauchy hypersurface $\Sigma \in {\cal{M}}$, on which
the $4$-dimensional regions ($C_1$,$C_2,{\cal{B}}$) become
$3$-dimensional surfaces. We still have a foliation of the buffer zone
by an infinite number of timelike junction hypersurfaces ${\cal{J}}_i$
of $\Sigma$. These hypersurfaces are now actually submanifolds with
boundary of co-dimension $1$ with respect to $\Sigma$ but co-dimension
$2$ with respect to ${\cal{M}}$.

In either $C_n$ we can now define the $3$-metric ($h_{ab}^{(n)}$) and
the extrinsic curvature ($K_{ab}^{(n)}$) of $\Sigma$ by projecting the
asymptotically matched approximate metrics $g_{\mu\nu}^{(n)}$ to this
hypersurface~\cite{Baumgarte:2002jm}. In analogy with the previous
section, let us then define these objects via
\ba
\label{joined-data}
h_{ab} &=& F(\ell_i) \; h_{ab}^{(1)} + \left[1 - F(\ell_i)\right] \; h_{ab}^{(2)}, 
\nonumber \\
K_{ab} &=& F(\ell_i) \; K_{ab}^{(1)} + \left[1 - F(\ell_i)\right] \; K_{ab}^{(2)}.
\ea
In \eref{joined-data}, $F(\cdot)$ is a proper transition function as
defined by \eref{asy-trans}. Provided this transition function
satisfies conditions $(ii)$ and $(iii)$ of Theorem~1, then
\eref{joined-data} satisfies the constraint equations of general
relativity.

There might be some concern that the extrinsic curvature of
\eref{joined-data} does not correspond to the same hypersurface
$\Sigma$ as that described by the $3$-metric because derivatives of
the transition functions have been neglected. However, if the
transition functions satisfy the conditions of Theorem~1, then these
derivatives are of the same order as the uncontrolled remainders. In
order to show this, we can compute the extrinsic curvature from the
spatial metric of \eref{joined-data} directly, {\textit{i.e.}}
\be
\label{sec-form-f}
K_{ab} = F(\ell_i) \; K_{ab}^{(1)} + \left[1 - F(\ell_i)\right] \;
K_{ab}^{(2)} - \frac{1}{2} \left\{h_{ab}\right\} {\cal{L}}_{\vec{n}}
F(\ell_i),  
\ee
where ${\cal{L}}_{\vec{n}}$ is the Lie derivative along the normal
vector to $\Sigma$. Clearly, since $F(\cdot)$ is a scalar function,
the Lie derivative reduces to the directional partial derivative of
this function along the normal vector, {\textit{i.e.}}
${\cal{L}}_{\vec{n}} = n^{\alpha} F_{,\alpha}$. Here $n^{\alpha}$
could be that associated with either of the approximate solutions,
since in the buffer zone $n^{\alpha}_{(1)} - n^{\alpha}_{(2)} =
{\cal{O}}(\epsilon_1^{\ell_1+1},\epsilon_2^{\ell_2+1})$.  Since the
$3$-metrics are already asymptotic to each other up to uncontrolled
remainders of ${\cal{O}}(\epsilon_1^{\ell_1+1},\epsilon_2^{\ell_2+1})$
by \eref{asympt-cond-deriv}, we must only require that the Lie
derivative be of ${\cal{O}}(\epsilon_1,\epsilon_2)$. This condition is
consistent with the conditions of the asymptotic junction theorem.

We see then that the global $3$-metric and extrinsic curvature of
\eref{joined-data} represent the same data as that obtained from
\eref{joined-g-asy} directly up to the uncontrolled remainders in the
approximations. We can then formulate the following theorem, which can
be viewed as a corollary of Theorem $1$:
\begin{theorem} 
  Consider a spacetime manifold ${\cal{M}}$ with approximate metrics
  $g^{(1)}_{\mu \nu}$ and $g^{(2)}_{\mu \nu}$ that satisfy the vacuum
  Einstein equations up to uncontrolled remainders of
  ${\cal{O}}(\epsilon_1^{\ell_1+1})$ and
  ${\cal{O}}(\epsilon_2^{\ell_2+1})$ on submanifolds with boundary $C_1$
  and $C_2$ respectively. Let these submanifolds intersect on a
  $4$-volume ${\cal{B}} = C_1 \cap C_2$ and foliate the
  $3$-dimensional projection of ${\cal{B}}$ onto a Cauchy hypersurface
  $\Sigma$ with timelike junction hypersurfaces ${\cal{J}}_i$.
  Consider the $3$-metric and extrinsic curvature of $\Sigma$
  constructed via
\ba
\label{eq:theorem:3}
h_{ab} &=& F(\ell_i) \; h_{ab}^{(1)} + \left[1 - F(\ell_i)\right] \; h_{ab}^{(2)}, 
\nonumber \\
K_{ab} &=& F(\ell_i) \; K_{ab}^{(1)} + \left[1 - F(\ell_i)\right] \; K_{ab}^{(2)}, 
\ea
where $F(\cdot)$ is a proper transition function as defined via
\eref{asy-trans}, $\ell_i$ is the proper distance to ${\cal{J}}_i$ on
$\Sigma$, and $\{h_{ab}^{(1,2)},K_{ab}^{(1,2)}\}$ are the $3$-metric
and extrinsic curvature of $\Sigma$ associated with $g_{\mu
  \nu}^{(1,2)}$.  Then, the $3$-metric and extrinsic curvature
\eref{eq:theorem:3} satisfy the constraint equations of General
Relativity on $\Sigma$ to the same order as $g_{\mu \nu}^{(1,2)}$ if
and only if the following conditions are satisfied:
\ba
&(i)& \qquad {\textrm{The metrics $g_{\mu\nu}^{(1,2)}$ have been
    asymptotically}}
\nonumber \\
&& \qquad \; {\textrm{matched in ${\cal{B}}$ up to uncontrolled
    remainders }}
\nonumber \\
&& \qquad \; {\textrm{of 
    ${\cal{O}}(\epsilon_1^{\ell_1+1},\epsilon_2^{\ell_2+1})$}},  
\nonumber \\
&(ii)& \qquad F_{,\alpha} = {\cal{O}}(\epsilon_1,\epsilon_2), \quad \textrm{in} \quad {\cal{B}}, 
\nonumber \\
&(iii)& \qquad F_{,\alpha\beta} = {\cal{O}}(\epsilon_1^{0},\epsilon_2^{0}),
\quad \textrm{in} \quad {\cal{B}}. 
\ea
\end{theorem}
\begin{proof}
  The proof of this theorem is established by projecting the
  $4$-metric onto $\Sigma$. The $3$-metric is given by
\be
h_{\alpha \beta} = g_{\alpha \beta} + n_{\alpha} n_{\beta},
\ee
where $n_{\alpha}$ is the covariant normal vector to $\Sigma$. Using
\eref{joined-g-asy} and the fact that any smooth tensor can be
decomposed with transition functions via \eref{split}, we rewrite the
spatial part of the $3$-metric as
\be
\label{first-form-final}
h_{ab} = h_{ab}^{(1)} F(\ell_i) + \left[ 1 - F(\ell_i)\right] h_{ab}^{(2)}.
\ee
The extrinsic curvature is given by
\be
\label{sec-form}
K_{ab} = {\cal{L}}_{\vec{n}} h_{ab}.
\ee
Inserting \eref{first-form-final} into \eref{sec-form} we can rewrite
the extrinsic curvature as given in \eref{sec-form-f}. However, since
the transition function satisfies $F_{,\alpha} =
{\cal{O}}(\epsilon_1,\epsilon_2)$, the last term of that equation can
be neglected and we obtain
\be
K_{ab} = K_{ab}^{(1)} F(\ell_i) + \left[1 - F(\ell_i) \right] K_{ab}^{(2)}.
\ee

Now, recall that the $4$-metric satisfies the vacuum Einstein
equations up to uncontrolled remainders of
$O(\epsilon_1^{\ell_1+1},\epsilon_2^{\ell_2+1})$ by Theorem~1.
Therefore, this metric also satisfies the constraint equations, since
these are related to the temporal components of the Einstein tensor
and the normal vector to $\Sigma$. Recall here that the normal can be
that associated with either approximate solution, since these vectors
are asymptotic to each other inside the buffer zone. Since the
$3$-metric and extrinsic curvature come directly from a projection of
this $4$-metric, it follows that this data must also satisfy the
constraints to the same order.
\end{proof}

\section{A simple example}
\label{calc-trans-func}
In this section, we investigate the impact of different transition
functions on the satisfaction of the Einstein equations. For this
purpose, we pick a metric that has already been matched
in~\cite{Yunes:2006iw}, henceforth paper~I. This metric represents a
binary system of Schwarzschild black holes in a quasicircular orbit.
We here show explicitly that as long as the transition functions
satisfy the conditions of Theorems~1 and~2, the joined metric
satisfies the vacuum Einstein equations and, thus, also the constraint
equations.

As explained in paper~I, the manifold can be divided into $3$ regions
(\tref{table}).
\begin{table}
\caption{\label{table}Description of the division of the spacetime
  into zones.}
\begin{indented}
\item[]\begin{tabular}{@{}cccc}
\br
Zone & $r_{in}$ & $r_{out}$  & $\epsilon_{n}$ \\
\mr
Inner zone BH $1$ (${\cal{C}}_1$) & 0 & $\ll b$ & $\bar{r}_1/b$ \\
Inner zone BH $2$ (${\cal{C}}_2$) & 0 & $\ll b$ & $\bar{r}_2/b$ \\
Near zone (${\cal{C}}_3$) & $\gg m_A$ & $\ll \lambda/2 \pi$ & $m_A/r_A$ \\
\br
\end{tabular}
\end{indented}
\end{table}
The symbols of \tref{table} represent the following quantities: $C_n$
labels the $n$th regions, where $n = \{1,2,3\}$; $A$ labels the black
hole, with $A = \{1,2\}$; $r_{in}$ and $r_{out}$ are the approximate
inner and outer boundary radius of each $C_n$ regions as measured from
the $A$th black hole; $r_A$ and $\bar{r}_A$ are the radial distances
as measured from the $A$th black hole in near and inner zone
coordinates; $\epsilon_{n}$ is the perturbation parameter used in the
approximate solution $g_{\mu\nu}^{(n)}$ in $C_n$; $m_A$ is the mass of
the $A$th black hole; $b$ is the orbital separation on $\Sigma$;
$\lambda$ is the gravitational wavelength. The quantity $b$ is usually
defined as the black hole separation in the near zone, which we shall
see coincides with the black hole separation in the inner zone up to
uncontrolled remainders, namely $b = \bar{b} +
{\cal{O}}(\epsilon_1,\epsilon_2)$.  Technically, there is a fourth
region beyond the near zone, but we neglect this here since it does
not affect the study of transition functions. For a detailed
description of this subdivision and a pictorial representation of
these zones refer to \cite{Yunes:2005nn,Yunes:2006iw}.

These regions overlap clearly in two buffer zones, which on the
$t=\bar{t}=0$ slice can be defined by the following inequalities:
$O_{13} = C_1 \cap C_3$ ($m_1 \ll r_1 \ll b$) and $O_{23} = C_2 \cap
C_3$ ($m_2 \ll r_2 \ll b$), where $r_A$ is the radial distance from
the $n$th black hole to a field point. Asymptotic matching and the
transition functions act in these buffer zones, which can only be
defined asymptotically and, thus, any statement regarding them must be
interpreted in that sense.  In particular, this implies that any
quantity that is valid in the buffer zone need not be valid near the
boundary of the buffer zone ({\textit{i.e.}}, as $r_A \to b$ or $r_A
\to m_A$.)  Also note that these buffer zones are actually $4$-volumes
and can be foliated by an infinite number of timelike junction
hypersurfaces.

Different approximations are used in each zone to solve the Einstein
equation. In either inner zone, black hole perturbation theory allows
us to obtain a tidally perturbed metric. Let us concentrate on inner
zone $1$ near black hole $1$, since the metric in the other inner zone
can be obtained via a symmetry transformation. In $C_1$, the perturbed
metric is given in isotropic corotating coordinates
$x^{\bar{a}}=\{\bar{t},\bar{x},\bar{y},\bar{z}\}$ by
\ba
\label{internalmetricICC}
g^{(1)}_{\bar{0}\bar{0}} &\approx& H_t + H_{s1} \Omega^2
\left(\bar{x}^2+\bar{y}^2\right) + 2 H_{st} \bar{x} {\frac{\Omega}{{b^2}}}
\left(\bar{x}^2+\bar{y}^2-\bar{z}^2\right), 
\nonumber \\
g^{(1)}_{\bar{0}\bar{a}} &\approx&  - H_{s1} \Omega
\epsilon_{\bar{a}\bar{b}\bar{3}} x^{\bar{b}} + \frac{H_{st}}{b^2}
\left[ \bar{y} \left(\delta_{\bar{a}}^{\bar{3}} \bar{z} -
    \delta_{\bar{a}}^{\bar{1}} \bar{x} \right) 
+ \left(\bar{x}^2 - \bar{z}^2\right) \delta_{\bar{a}}^{\bar{2}}
\right],  
\nonumber \\ 
g^{(1)}_{\bar{a}\bar{b}} &\approx& \delta_{\bar{a}\bar{b}} H_{s1} -
H_{s2} \frac{x^{\bar{a}} x^{\bar{b}}}{b^2},  
\ea
where $\Omega$ is the angular velocity of the perturbation,
$\epsilon_{\bar{a}\bar{b}\bar{c}}$ is the standard Levi-Civita symbol
with convention $\epsilon_{\bar{1}\bar{2}\bar{3}} = 1$ and where
$\delta^{\bar{a}}_{\bar{b}}$ is the Kronecker delta. In
\eref{internalmetricICC} we used the shorthand
\ba
H_{st} &=& 2 m_2 \sqrt{\frac{m}{b^3}} 
\left(1 - {\frac{M_1}{2 {\bar{r}}}}\right)^2 \left(1 + {\frac{M_1}{2 {\bar{r}}}} \right)^4,
\nonumber \\ 
H_{s1} &=& \left(1 + {\frac{M_1}{2 {\bar{r}}}}\right)^4 \left\{1 + 2{\frac{m_2}{b^3}} 
{\bar{r}}^2 P_2\left(\frac{\bar{x}}{{\bar{r}}}\right)
 \left[\left(1 + {\frac{M_1}{2 {\bar{r}}}}\right)^4 - 2
  {\frac{M_1^2}{{\bar{r}}^2}}\right]\right\},  
\nonumber \\
H_{s2} &=& \left(1+\frac{M_1}{2{\bar{r}}}\right)^4
\left(1+{\frac{M_1^2}{4{\bar{r}}^2}}\right) {\frac{4m_2 M_1}{b {\bar{r}}}}  
P_2\left(\frac{\bar{x}}{{\bar{r}}}\right) ,
\nonumber \\ 
H_{t} &=& - \left(\frac{1-M_1/2 {\bar{r}}}{1+M_1/2 {\bar{r}}}\right)^2 
 + 2 \left(1-\frac{M_1}{2 {\bar{r}}}\right)^4 {\frac{m_2}{b^3}} {\bar{r}}^2
P_2\left(\frac{\bar{x}}{{\bar{r}}}\right),
\ea
where $M_1$ is the mass of the background black hole, $m_2$ is the
mass of the binary companion that is causing the perturbation, $m =
M_1 + m_2$ is the total mass, $\bar{r} = (\bar{x}^2 + \bar{y}^2 +
\bar{z}^2)^{1/2}$ and $P_2(\cdot)$ is the second Legendre polynomial.
\Eref{internalmetricICC} then satisfies the linearized Einstein
equations in $C_1$. These equations are identical to~($18$) and~($19$)
of paper~I and solve the Einstein equations up to uncontrolled
remainders of ${\cal{O}}(\bar{r}_1/b)^3$. We refer the reader to that
paper for an explanation of the derivation of this metric.

In the near zone, a post-Minkowskian expansion is used to find an
approximate solution. This solution in corotating ADMTT coordinates
$x^{a}=\{t,x,y,z\}$ [~($7$)-($12$) in paper~I] is given by
\ba
\label{admtt-metric}
g_{ab}^{(3)} &=& \Psi^4 \delta_{ab} ,
\nonumber \\
g_{0a}^{(3)} &=& g_{ab}^{(3)} \beta^{(3) b},
\nonumber \\
g_{00}^{(3)} &=& g_{0a}^{(3)} \beta^{(3) a} - (\alpha^{(3)})^2,
\nonumber \\
\ea
where we introduced a post-Newtonian conformal factor
\be
\label{Psi_ADMTT}
\Psi = 1 + \frac{m_1}{2 r_1} + \frac{m_2}{2 r_2},
\ee
and where the post-Newtonian lapse and shift are given by
\ba
\label{alpha_ADMTT}
\alpha^{(3)} &=& \frac{2-\Psi}{\Psi}
\nonumber \\
\beta^{(3) i} &=& 
\frac{m_1}{r_1} \left[\frac{1}{2}
\left(v_{1}^{i} - \vec{v}_1 \cdot \vec{n}_1 n_{1}^{i} \right) 
- 4 v_{1}^{i} \right]  
\nonumber \\
&+&
\frac{m_2}{r_2} \left[\frac{1}{2}
\left(v_{2}^{i} - \vec{v}_2 \cdot \vec{n}_2 n_{2}^{i} \right) 
- 4 v_{2}^{i} \right] - \epsilon_{ik3} \omega x^{k}.
\ea
In these equations, the radial distance to the $A$th black hole is
given by $r_A = (x_A^2 + y^2 + x^2)^{1/2}$, where $x_1 = x - m_2 b/m$
and $x_2 = x + m_1 b/m$. These equations solve the Einstein equations
in $C_3$ up to uncontrolled remainders of ${\cal{O}}(m_A/r_A)^2$.
Once more, we refer the reader to that paper for an explanation of the
derivation of such a metric. In these equations, the non-zero
components of the velocities $\vec{v}_A$ and the unit vectors
$\vec{n}_A$ are given in the $t=0$ slice by
\ba
v_{1}^{2} &=&  \omega \frac{m_2}{m} b ,\qquad
v_{2}^{2}  =  -\omega \frac{m_1}{m} b ,\qquad
\nonumber \\
n^{k}_{A} &=& \frac{x^{k} - \xi^{k}_{A}}{r_A} ,\ \ \
\xi^{1}_1 =  \frac{m_2}{m} b , \ \
\xi^{1}_2 = -\frac{m_1}{m} b,\ \
\ea
where $w$ is the post-Newtonian angular velocity given by 
\be 
\label{ang-vel-def}
\omega=\sqrt{\frac{m}{b^3}}\left[1+ \frac{1}{2} 
       \left(\frac{\mu}{m}-3\right) \frac{m}{b}\right],
\ee
with errors of ${\cal{O}}(m/b)^{5/2}$. \Eref{ang-vel-def} appears also
in paper~I and in~(60) of~\cite{Tichy:2002ec}, where $\mu = m_1 m_2/m$
is the reduced mass of the system.

Asymptotic matching was performed in paper~I and a coordinate and
parameter transformation was found such that \eref{asympt-cond} is
satisfied. For buffer zone~1, $O_{13}$, such a transformation to
${\cal{O}}[m_A/r_A,(\bar{r}_A/b)^2]$ is given by
\ba
\label{matching-conds}
x^{\bar{\mu}} &=& x^{\sigma} \left[\delta_{\sigma}{}^{\bar{\mu}} + 
  \eta_{\sigma}{}^{\bar{\mu}} \frac{m_2}{b} \left( 1 -
    \frac{x}{b}\right) \right] + \delta_{y}{}^{\bar{\mu}} t
\frac{m_2}{\sqrt{m b}},
\nonumber \\
M_A &=& m_A, \qquad \Omega = \omega,
\ea
where $\eta_{\mu \nu}$ is the Minkowski metric and $\delta_{\mu \nu}$
is the $4$-Euclidean metric. Note that $x^{\bar{\mu}} = x^{\mu}$ to
zeroth order as previously mentioned and, when describing certain
figures, we will sometimes use them interchangeably.  One might worry
that \eref{matching-conds} is not bounded above, since
$x^{\bar{\sigma}} \to \infty$ as $t \to \infty$. However, recall that
the buffer zone is technically a $4$-volume delimited by the
boundaries of the regions of validity of the approximations. In this
sense, time cannot go to infinity, because then the post-Newtonian
metric would break down as $t$ approaches the time of coalescence.
Furthermore, when constructing initial data, $t = 0$ and $\bar{t} =
0$, and it is clear that $t \to \infty$ is not allowed.  In paper~I,
asymptotic matching is carried out to slightly higher order, but since
the purpose of this section is to study transition functions we use
\eref{matching-conds} instead.  The metric in inner zone $2$ and the
matching coordinate and parameter transformations in buffer zone~2,
${\cal{O}}_{23}$, can be obtained by applying the following
transformation to \eref{internalmetricICC} and \eref{matching-conds}:
\ba
\label{sym-transf}
1 \to 2, \qquad x \to -x, \qquad y \to -y, \qquad z \to z.
\ea

In order to study these transition functions, we pick a particular
physical system. We choose a system of equal mass black holes $m_1 =
m_2 = 0.5 m$ separated by an orbital distance $b = 20 m$. The black holes
are non-spinning and located on the $x$-axis at $x = 10 m$ and $x =
-10 m$ for the first and second black hole respectively. The black holes are
orbiting in the counter-clockwise direction about the $z$-axis. The
buffer zones are then given by $O_{A3}: 20 \gg r_A/m \gg 0.5$. This
system is used to study different transition functions, but we should
note that the asymptotically matched metric of paper~I is valid for a
wide range of systems.

We here keep the physical system fixed and pick several different
transition functions to investigate how these functions change the
satisfaction of the Einstein equations. We consider the following type
of transition functions:
\begin{itemize}
\item {\bf{Transition function $1$ (TF~1)}}
\be
\fl
\label{trans1}
f_1(r) = \cases{
0  &, $r \leq r_-$\\
\frac{1}{2} \left\{ 1 + \tanh \left[ \frac{s}{\pi} \left(
\tan(\frac{2 m \pi}{b w}(r-r_-)) 
\right. \right. \right. & \\ \left. \left. \left. 
\qquad \; - \frac{q^2}{\tan(\frac{2 m \pi}{b
    w}(r-r_-))}\right) \right] \right\}  &, $r_- < r < r_+$ \\
1 &, $r \geq r_+$,}
\ee
where $r_-$ and $r_+$ determine where the transition begins and ends,
$w$ determines the size of the region, and $s$ and $q$ determine the
slope of the transition function roughly when $f_1 \approx 1/2$. This
function is similar to that used in paper~I and we use similar
parameters, {\textit{i.e.}}
\ba
\label{transpars1}
r_- &=& m, \quad
r_+ = b - m, 
\nonumber \\
w   &=& 9 m, \quad
q   = 0.2, \quad
s   = b/m.
\ea

\item {\bf{Transition function 2 (TF~2) and 3 (TF~3)}}
\be
\label{trans2}
f_2(r) = \exp\left({\frac{b}{2m}\frac{r - d}{w}}\right) \left[1 +
  \exp\left({\frac{b}{2m}\frac{r - d}{w}}\right)\right]^{-1}, 
\ee
where $w$ is approximately the size of the transition window and $d$
is approximately the distance from the $A$th black hole at which the
derivative of the transition function peaks. For TF~2 we choose the
following parameters
\be
w = 7 m, \qquad d = 6 m,
\ee
while for TF~3 we choose
\be
w=m, \qquad d=8 m. 
\ee
\end{itemize}
For all these transition functions, the transition window $w$ is
defined approximately as the size of the region inside which the
function ranges from $0.01$ to $0.99$. 

The purpose of this paper is not to perform a systematic study of the
properties of transition functions, but to illustrate the theorems
discussed in earlier sections with a practical example. In fact, as
discussed in the introduction, transition functions were first
introduced in~\cite{Yunes:2005nn,Yunes:2006iw}, but the properties and
conditions that these functions must satisfy were not explored.  In
this paper, we are in essence following up on previous work and
providing further details that might be relevant to future
investigations. In particular, we shall investigate how the gluing of
asymptotically matched approximate solutions breaks down if improper
transition functions are chosen.

The transition functions that we have chosen clearly have different
properties. First, notice that the region where functions are
significantly different from zero or unity varies, because the size of
their respective transition windows $w$ is different. In particular,
TF~3 has the smallest transition window, followed by TF~2 and then
TF~1. Second, notice that the transition functions become roughly
equal to $1/2$ at different radii: $r \approx 7 m$ for TF~1, $r
\approx 6 m$ for TF~2 and $r \approx 8 m$ for TF~3. Third, notice that
the speed at which the transition functions tend to zero and unity is
also different. In particular, note that TF~3 tends to zero the
fastest, followed by TF~1 and finally TF~3. This speed is important in
the definition of proper transition functions in \eref{asy-trans},
because it prevents contamination of possibly divergent uncontrolled
remainders outside the buffer zone. Finally, notice that not all
functions are analytic, since TF~1 does not have a Taylor expansion
about $r = r_{\pm}$ (recall, however, that analyticity was not
required.)

\begin{figure}
\begin{center}
\includegraphics[scale=0.4,clip=true]{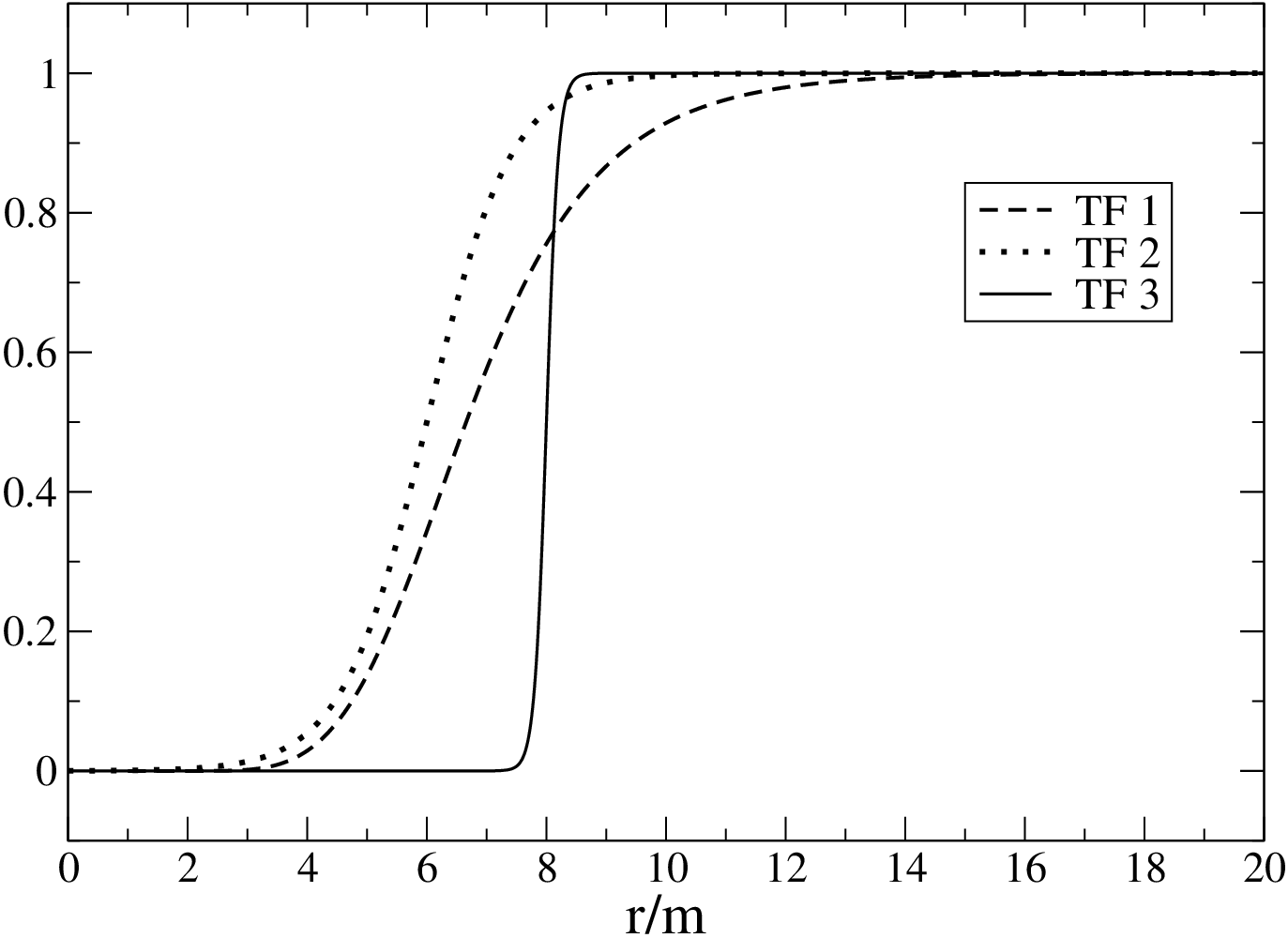}
\caption{\label{transfunc} This figure shows the first (TF~1), second
  (TF~2) and third (TF~3) transition functions as a dashed, dotted and
  solid lines respectively. Observe that these functions transition at
  different radii and at different speeds.}
\end{center}
\end{figure}

\begin{figure}
\begin{center}
\includegraphics[scale=0.35,clip=true]{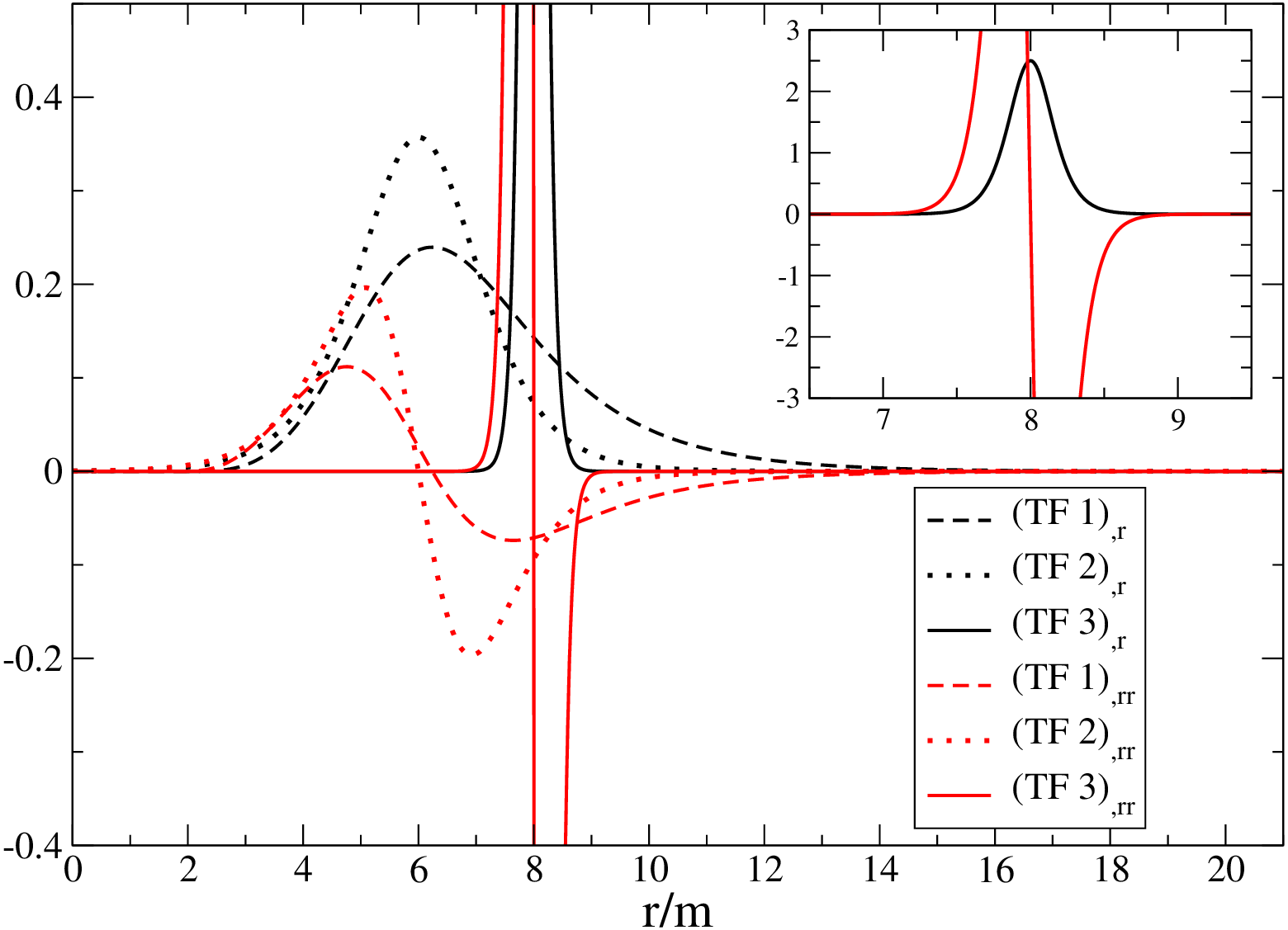}
\caption{\label{transfunc-deriv} This figure shows the first and second
  derivatives of the transition functions. The first derivative of the
  first [$(TF~1)_{,r}$], second [$(TF~2)_{,r}$] and third
  [$(TF~3)_{,r}$] transition functions are denoted with a dashed,
  dotted and solid black line respectively. The second derivative of
  the first [$(TF~1)_{,rr}$], second [$(TF~2)_{,rr}$] and third
  [$(TF~3)_{,rr}$] transition functions are denoted with a dashed,
  dotted and solid cyan (or light gray on black and white print) line
  respectively. In the inset we zoomed out to show better the
  derivatives of the third transition function.}
\end{center}
\end{figure}

These different features of the transition functions can be observed
in Figures~\ref{transfunc} and~\ref{transfunc-deriv}. \Fref{transfunc}
shows the behaviour of the different transition functions as a
function of radius, while~\fref{transfunc-deriv} shows their first
and second derivatives.
Observe in~\fref{transfunc} that the transition occurs at different
radii and that they transition at different speeds.  In
~\fref{transfunc-deriv} we can observe the difference in the
transitioning speed better. We can clearly see in this figure that the
derivatives of TF~1 are the smallest, followed by TF~2 and then TF~3.
However, both TF~1 and TF~2 have derivatives that are consistently of
order much less than ${\cal{O}}(1)$, while TF~3 has first derivatives
of ${\cal{O}}(1)$ or larger. The inset in this figure shows the large
derivatives of TF~3. Also note that the second derivatives of the
first and second transition functions are consistently smaller than
their first derivatives. This is a consequence of the size of the
transition window. If such a window were chosen to be smaller, as in
the case of TF~3, then the second derivative would become larger than
the first.

With these transition functions we can now construct pure joined
solutions via
\be
\label{joined-g-2BHS}
g_{\mu \nu} = f(\bar{r}_1) f(\bar{r}_2) g_{\mu \nu}^{(3)} + \left[1 - 
f(\bar{r}_1)\right] g_{\mu\nu}^{(1)} + \left[1 - f(\bar{r}_2) \right]
g_{\mu \nu}^{(2)},  
\ee
where we assume $g_{\mu\nu}^{(1,2)}$ has been transformed with the
coordinate and parameter transformation of \eref{matching-conds} and
where $f(\cdot)$ can be any of TF~1, TF~2 or TF~3.
\Eref{joined-g-2BHS} is an extension of \eref{joined-g-asy} applicable
to two buffer zones. Extensions to more than $2$ buffer zones are also
straightforward. 

\begin{figure}
\begin{center}
\includegraphics[scale=0.4,clip=true]{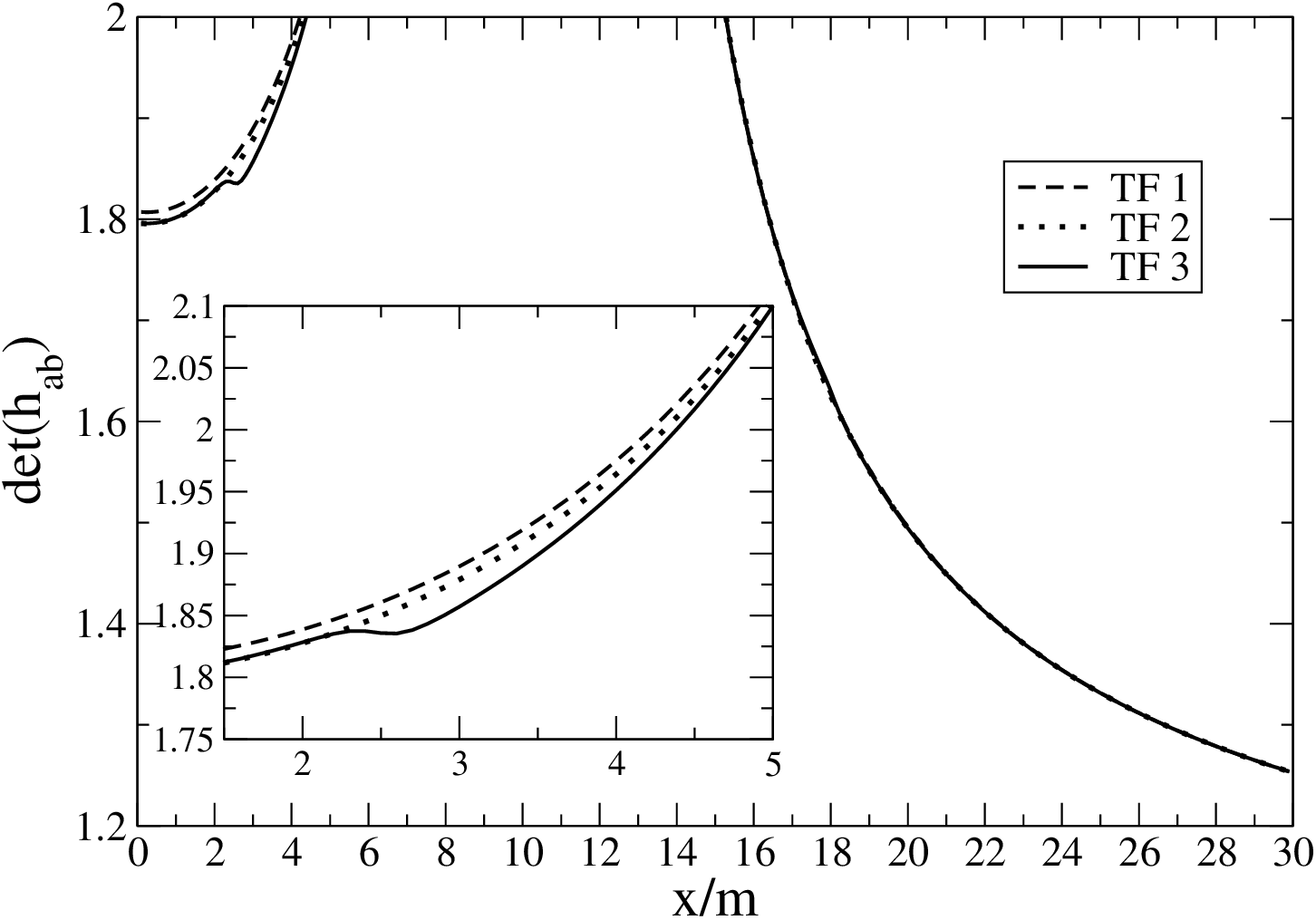}
\caption{\label{gTF} This figure shows the determinant of the spatial
  metric along the $x$-axis with TF~1 (dashed line), TF~2 (dotted
  line) and TF~3 (solid line.) Observe that the differences in the
  global metrics are small if at all visible. In the inset we zoom to
  the region where the curves look the most different.}
\end{center}
\end{figure}

Whether the joined metric of \eref{joined-g-2BHS} satisfies the
Einstein equations to the same order as the approximate solution
depends on the transition function used. We expect the metric
constructed with TF~1 to generate small violations because it is a
proper transition function that clearly satisfies the
differentiability conditions required in Theorems~1 and~2. On the
other hand, TF~2 and~3 do not satisfy these conditions because TF~2 is
not a proper transition function and TF~3 violates condition $(ii)$
and $(iii)$ of Theorem~1. This behaviour, however, is not clear by
simply looking at metric components. In \fref{gTF} we plot the
determinant of the spatial metric along the $x$-axis, corresponding to
the $x$ harmonic (near zone) coordinate, with $t=0$, $y=0$ and $z=0$.
This axis is where joined metrics are glued together with different
transition functions.  This determinant gives a measure of the volume
element on $\Sigma$ and, as one can see in the figure, the difference
when different transition function are used is small. This behaviour
is not unique to the $x$-axis, but is actually observed along the
other axis as well.  When this is the case, we only show the behaviour
along the $x$-axis in order to avoid redundancy.

\begin{figure}
\begin{center}
\includegraphics[scale=0.4,clip=true]{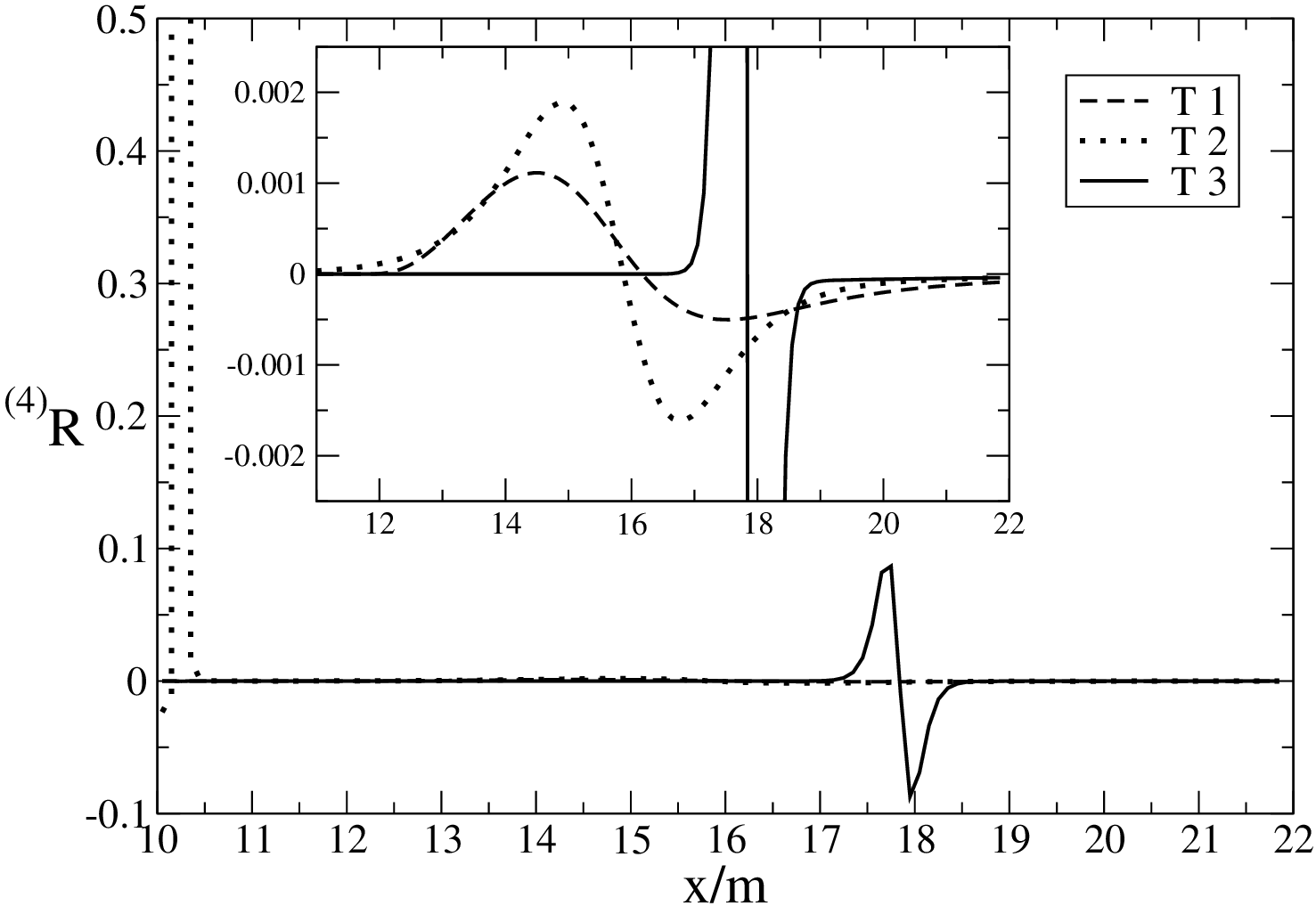} 
\caption{\label{R4} $4$-Ricci scalar along the x-axis for the 
  global metrics joined with TF~1 (dashed line), TF~2 (dotted line)
  and TF~3 (solid line.) Observe that the $4$-Ricci scalar is small
  everywhere for the metric constructed with TF~1, but it has spikes
  for that built with TF~2 and~3. The inset zooms to the region $x \in
  [11,22]$ so that the difference between the $4$-Ricci scalar
  calculated with TF~1 and~2 are more noticeable.}
\end{center}
\end{figure}

Although the volume element computed with different transition
functions is similar, metrics joined with different transition
functions do not satisfy the Einstein equations to the same order.  In
~\fref{R4}, we plot the $4$-Ricci scalar calculated with the global
metrics joined with different transition functions. This plot is
representative of the behaviour of the $4$-Ricci scalar in the entire
domain, although here we plot it only along the $x$-axis. We also plot
only the region $x/m > 10$, to the right of where BH~1 is located at
$t=0$, because the behaviour of the $4$-Ricci scalar is symmetric
about $x=0$ and the transition functions are symmetric about
$\bar{r}_A = 0$. Thus, the behaviour of the $4$-Ricci scalar in other
regions of the domain is similar to that shown in~\fref{R4} (see
paper~I for contour plots of some of these quantities.)

There are several features in~\fref{R4} that we should comment on.
First, observe that the $^{(4)}R$ is everywhere smaller than the
uncontrolled remainder in the approximate solutions (roughly
${\cal{O}}(m/b)^2 \approx 0.0025$ in the buffer zone for the system
considered) when the joined metric is constructed with TF~1. However,
this is not the case for the metrics constructed with TF~2 and~3. For
those metrics, $^{(4)}R$ has spikes close to $r_A = 0$ ($x \approx 10
m$) and $r_A = 8 m$ ($x \approx 18 m$) for TF~2 and~3 respectively.
The spike resulting from the metric constructed with TF~3 is
associated with the small size of its transition window, which forces
large derivatives in the transition function, thus violating
conditions $(ii)$ and $(iii)$ of Theorems~1 and~2. The spike in the
metric built with TF~2 is related to the fact that this function does
not tend to zero faster than the uncontrolled remainders of the
post-Newtonian near zone metric close to BH~A. As we discussed
earlier, the rate at which transition functions tend to zero and unity
is important to avoid contamination from divergent uncontrolled
remainders in the approximations. In this case, TF~2 is of
${\cal{O}}(m/b)^3$ as $r/m \to m/b$, while the post-Newtonian Ricci
scalar diverges as $m/r^3$ as $r \to 0$ and is in fact of
${\cal{O}}(b/m)^3$ as $r/m \to m/b$.  Since the transition function is
not able to eliminate this contamination from the near zone metric,
the Einstein equations are violated near either black hole for the
metric constructed with TF~2. From a physical standpoint, the failure
to use appropriate transition functions to create a pure joined
solution introduces a modification in the matter-sector of the
Einstein equations, namely the artificial creation of a shell of
matter. One can straightforwardly see from \eref{Tab}
and~\eref{sec-form-f} that the stress-energy tensor of this shell
depends on the non-vanishing derivatives of the transition function.

A deeper analysis of the spike in $^{(4)}R$ due to the joined metric
with TF~2 reveals that the conditions found are sufficient but
formally not necessary. The $4$-Ricci scalar constructed with the
post-Newtonian metric ($^{(4)}R_{PN}$) diverges close to BH~A, since
this approximation breaks down in that region. Therefore, if the
transition function does not vanish identically, or faster than the
divergence in $^{(4)}R_{PN}$, then $^{(4)}R$ will present a spike.
This spike, however, would disappear if the transition function
decayed to zero faster than the divergence in $^{(4)}R_{PN}$.  In this
sense, eventhough the conditions discussed here are not necessary
since the definition of a proper transition function could be
weakened, they are certainly sufficient and universal.

The second feature we should comment on is the behaviour of $^{(4)}R$
in the region $x/m \in [11,17]$. As one can observe in the inset, the
transition functions indeed introduce some error in this region.
However, note that this error for the metric constructed with TF~1 is
smaller than the uncontrolled remainders in the approximate solutions.
Also note that this error is independent of the location of the center
of the transition window. In other words, asymptotic matching is
performed inside of a buffer zone and not on a patching surface, which
means that approximate solutions can be glued with transition
functions anywhere in the buffer zone away from the boundaries and not
just at a specific $2$-surface.  As one can see from the figure, the
metric constructed with TF~1 has the smallest error in this region,
followed by that built with TF~2 and~3.  This fact is not too
surprising because, as shown in the proof of Theorem~1, the error
introduced by the transition functions in the calculation of the Ricci
tensor scales with the first and second derivatives of these
functions. Finally, note that the error introduced by these functions
seems to be correlated to both the size and functional form of the
second derivative of the transition functions, as one can see by
comparing the inset of \fref{transfunc-deriv} to \fref{R4}. The reason
for such similarity, however, is beyond the scope of this paper.

Once an approximate global $4$-metric has been found, we can use it to
construct initial data. These data could be given for example by 
\ba
\label{joined-data-2BHS}
h_{ab} &=& f(\bar{r}_1) f(\bar{r}_2) h_{ab}^{(3)} + \left[1 - 
f(\bar{r}_1)\right] h_{ab}^{(1)} 
 + \left[1 - f(\bar{r}_2) \right] h_{ab}^{(2)}, 
\nonumber \\ 
K_{ab} &=& f(\bar{r}_1) f(\bar{r}_2) K_{ab}^{(3)} + \left[1 - 
f(\bar{r}_1)\right] K_{ab}^{(1)} 
 + \left[1 - f(\bar{r}_2) \right] K_{ab}^{(2)}. 
\ea
\Eref{joined-data-2BHS} is simply a generalization of
\eref{joined-data} for $2$ buffer zones. The extrinsic curvatures for
the inner zones and near zone are provided explicitly in paper~I.

Are we allowed to construct data with transition functions in this
way? There might be some doubt as to whether this is valid, since the
fact that the joined metric satisfies the Einstein equations does not
necessarily guarantee that we can use transition function to construct
the data itself. In particular, there might be some worry that the
extrinsic curvature calculated from \eref{joined-g-2BHS} generally
contains derivatives of the transition functions that
\eref{joined-data-2BHS} neglects. Theorem~2, however, ensures that
this construction is indeed valid, provided the transition functions
satisfy the same differentiability conditions proposed in Theorem~1.
This is because the derivatives of these functions are then small and,
in particular, the terms that are proportional to them are of the same
order as the uncontrolled remainders in the approximations.

\begin{figure}
\begin{center}
\includegraphics[scale=0.4,clip=true]{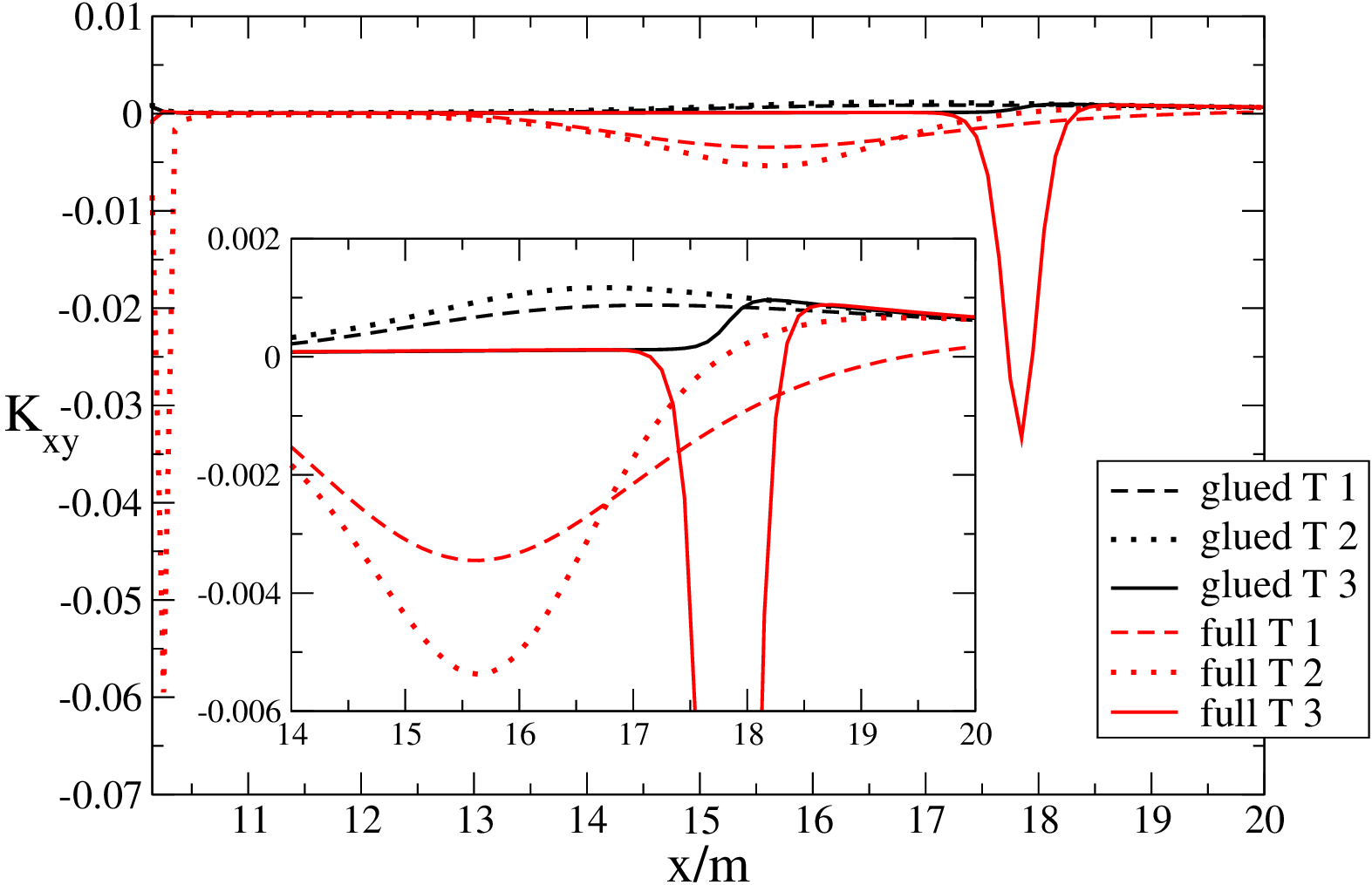} 
\caption{\label{K} $xy$-component of the extrinsic curvature along
  the $x$-axis, constructed via \eref{joined-data-2BHS} (black) for
  TF~1 (dashed line), TF~2 (dotted line) and TF~3 (solid line.) We
  also plot this component constructed by direct differentiation of
  \eref{joined-g-2BHS} (cyan, or light gray on black and white print)
  for TF~1 (dashed line), TF~2 (dotted line) and TF~3 (solid line.)
  The inset zooms to the region $x \in \{15,22\}$ so that the
  differences are more noticeable.}
\end{center}
\end{figure}

These expectations can be verified in~\fref{K}, where we plot the
$xy$-component of the extrinsic curvature along the $x$-axis
constructed both via \eref{joined-data-2BHS} (referred to as glued
T~1,~2 and~3 in the figure) and by direct differentiation of
\eref{joined-g-2BHS} (referred to as full T~1,~2 and~3 in the figure.)
We plot only the $xy$-component because this is the dominant term of
this tensor along the $x$-axis for the system considered and it shows
the main differences in using different transition functions. We have
checked that other quantities, like the trace of the extrinsic
curvature, behaves similarly. For contour plots of the extrinsic
curvature refer to paper~I.

As in the case of the Ricci scalar, there are several features of
\fref{K} that we should discuss. First, observe that that the
extrinsic curvatures constructed with TF~1 agree in the buffer zone up
to uncontrolled remainders. These remainders are of
${\cal{O}}(m/b)^{3/2} \approx 0.01$ in the buffer zone, because here
we have used a coordinate transformation from the matching scheme that
is valid only up to ${\cal{O}}(m/b)$. The inset zooms to a region
close to the outer boundary of the buffer zone in order to show this
agreement better.  The humps in this region are produced by the
non-vanishing Lie derivatives of the transition functions. Observe
that, as expected, these humps are smallest for TF~1, followed by TF~2
and~3.  Our analysis suggests that if we performed matching to higher
order, the agreement would be better and the size of the humps would
decrease.  The agreement is good for the curvatures constructed with
TF~1, but the curvatures built with TF~2 and~3 have strong spikes
roughly near the boundary of the buffer zone.  These spikes arise
because TF~2 violates the definition of a proper transition function,
while TF~3 violates conditions $(ii)$ and $(iii)$ of Theorem~2.

We have then seen in this section how Theorems~1 and~2 can aid us in
constructing transition functions.  Even though we have not shown the
error in the constraints, we have checked that this error presents the
same features as those shown in \fref{R4}. We have also seen the
importance of restricting the family of transition functions to those
that satisfy conditions $(ii)$ and $(iii)$ of Theorems~1 and~2, as
well as the definition of a proper transition function. The
construction of pure, or mixed, joined solutions can then be carried
out with ease as long as transitions functions are chosen that satisfy
the conditions suggested here.

\section{Conclusion}
\label{conclusion}
We studied the construction of joined solutions via transition
functions. In particular, we focused on pure joined solutions,
constructed by gluing analytical approximate solutions inside some
buffer zone where they were both valid. The gluing process was
accomplished via a weighted-linear combination of approximate
solutions with certain transition functions. We constrained the family
of allowed transition functions by imposing certain sufficient
conditions that guarantee that the pure joined solution satisfies the
Einstein equations to the same order as the approximations. With these
conditions, we formulated and proved a theorem that ensures that the
joined solution is indeed an approximate solution to the Einstein
equations. We extended these conditions to projections of the joined
solution onto a Cauchy hypersurface. We verified that the data on this
hypersurface can itself be constructed as a weighted-linear
combination with the same transition functions as those used for the
joined solution. We proved that if these functions satisfy the same
sufficient conditions, the data is guaranteed to solve the constraints
of the Einstein equations to the same order as the approximations.

We explicitly verified these theorems numerically by considering a
binary system of non-spinning black holes. The approximate solutions
used were a post-Newtonian expansion in the far field and a perturbed
Schwarzschild solution close to the black holes. We considered three
different kinds of transition functions, two of which violated the
conditions proposed in the theorems. The joined solutions constructed
with these transition functions were shown to produce large violations
to the Einstein equations. These violations were interpreted as the
introduction of a matter-shell, whose stress-energy tensor was shown
to be related to derivatives of the transition functions. The joined
solution constructed with the transition function that did satisfy the
conditions of the theorems was seen to introduce error comparable to
that already contained in the approximate solutions. We further
verified that projections of this joined solution to a Cauchy
hypersurface can also be constructed as a weighted-linear combination
of projections of approximate solutions with transition function.
These joined projected solutions were seen to still satisfy the
constraints of the theory to the same order as the approximate
solutions provided the transition functions satisfied the conditions
of the theorems.

The glue proposed here to join approximate solutions has several
applications to different areas of relativistic gravitation. Initial
data for numerical simulations could, for example, be constructed once
approximate solutions to the system have been found and asymptotically
matched inside some buffer zone. In particular, initial data for a
binary system of spinning black holes could be generated by
asymptotically matching and gluing a tidally perturbed Kerr
metric~\cite{Yunes:2005ve} to a post-Newtonian expansion (see
~\cite{Owen}.)  One could also study the absorption of energy and
angular momentum~\cite{Poisson:2004cw}, as well as the motion of test
particles~\cite{Hughes:2001jr}, in the spacetime described by such
pure joined solutions. Finally, one could extend the scheme developed
here to formalize the construction of mixed joined solutions. Such
solutions could, for example, be composed of a post-Newtonian metric
glued to a numerically simulated metric or a semi-analytical
approximate
metric~\cite{Buonanno:2000ef,Buonanno:2005xu,Sopuerta:2006wj,Sopuerta:2006et}.
Such mixed joined solutions could then be used to construct reliable
waveform templates for extremely non-linear physical scenarios.

\ack
  
We would like to thank Ben Owen for his continuous support and
encouragement. We would also like to thank Martin Bojowald, Victor
Taveras, Carlos Sopuerta and Ben Owen for reading this manuscript and
providing useful comments. Finally, we would like to acknowledge the
support of the Institute for Gravitational Physics and Geometry and
the Center for Gravitational Wave Physics, funded by the National
Science Foundation under Cooperative Agreement PHY-01-14375. This work
was also supported by NSF grants PHY-02-18750, PHY-02-44788,
PHY-02-45649, PHY-05-55628.

\section*{References}
 
\bibliographystyle{unsrt} 
\bibliography{junction.bib}

\end{document}